\documentclass[pra,twocolumn]{revtex4-1}
\usepackage{graphicx,bm,color}

\newcommand{\Red}{\color{red}}

\newcommand{\bfk}{{\bf k}}
\newcommand{\bfB}{{\bf B}}
\newcommand{\ua}{\uparrow}
\newcommand{\da}{\downarrow}

\newcommand{\bfr}{{\bf r}}
\newcommand{\bfq}{{\bf q}}
\newcommand{\bfv}{{\bf v}}

\newcommand{\si}{\sigma}
\newcommand{\bsi}{{\bar{\sigma}}}

\newcommand{\h}{\hat{H}}
\newcommand{\hbS}{\hat{\bf S}}
\newcommand{\hbJ}{\hat{\bf J}}

\newcommand{\hbr}{\hat{\bf r}}

\newcommand{\bfb}{{\bf b}}
\newcommand{\bfA}{{\bf A}}
\newcommand{\bfE}{{\bf E}}
\newcommand{\bfQ}{{\bf Q}}
\newcommand{\hbp}{\hat{\bf p}}
\newcommand{\hp}{\hat{p}}
\newcommand{\hbL}{\hat{\bf L}}
\newcommand{\bfz}{{\bf e}_z}
\newcommand{\bfx}{{\bf e}_x}

\newcommand{\SW}{spin wave }
\newcommand{\SWd}{spin-wave }

\newcommand{\sw}{_{\rm sw}}
\newcommand{\sws}{S_{\rm sw}} 

\begin{document}

\title{Chirality and intrinsic dissipation of spin modes in two-dimensional electron liquids}

\author{Irene D'Amico} \thanks{The authors contributed equally to this review; their names are listed in alphabetic order.}
\address{Department of Physics, University of York, York YO10 5DD, United Kingdom}

\author{Florent Perez}
\address{Institut des Nanosciences de Paris, CNRS/Universit\'e Paris VI, Paris 75005, France}

\author{Carsten A. Ullrich}
\address{Department of Physics and Astronomy, University of Missouri, Columbia, MO 65211, USA}

\begin{abstract}
We review recent theoretical and experimental developments concerning collective spin excitations in two-dimensional electron liquid (2DEL) systems, with particular emphasis on the interplay between many-body and spin-orbit effects, as well as the intrinsic dissipation due to the spin-Coulomb drag.
Historically, the experimental realization of 2DELs in silicon inversion layers in the 60s and 70s created unprecedented opportunities to probe
subtle quantum effects, culminating in the discovery of the quantum Hall effect. In the following years,
high quality 2DELs were obtained in doped quantum wells made in typical semiconductors like GaAs or CdTe. These systems became important test beds for quantum many-body effects due to Coulomb interaction, spin dynamics, spin-orbit coupling, effects of applied magnetic fields, as well as dissipation mechanisms.
Here we focus on recent results involving chiral effects and intrinsic dissipation of collective spin modes: these are not only of fundamental interest but also important towards demonstrating new concepts in spintronics. Moreover, new realizations of 2DELs are emerging beyond traditional semiconductors, for instance in multilayer graphene, oxide interfaces, dichalcogenide monolayers, and many more. The concepts discussed in this review will be relevant also for these emerging systems.
\end{abstract}



\maketitle


\section{Introduction} \label{sec:1}
Understanding and controlling the mechanisms that create or dissipate collective spin motion is, on the one hand, of fundamental interest. On the other hand, this topic has also drawn considerable attention since the spin degrees of freedom are expected to overcome some of the well-known limitations of charge-based electronics \cite{Chumak2015,Zutic2004}. The spins of charged particles interact through the Coulomb-exchange interaction and can precess coherently in a collective motion in the form of a spin wave \cite{SpinWavesBook,AuerbachBook}.

Spin waves propagate and can carry spin-based information over a significant distance that depends on the product of the propagation velocity (which grows with the strength of the Coulomb exchange) and the lifetime (which is inversely proportional to the strength of dissipative mechanisms that destroy the coherent motion).
Understanding the intrinsic laws which determine the balance between these opposite trends is a fundamental topic which has been addressed in various condensed matter systems,
including insulating ferromagnets \cite{Kajiwara2010}, conducting ferromagnets \cite{Zakeri2014,Stigloher2016}, bilayer systems \cite{Plaut1997,Tifrea2002,Bolcatto2003,Abedinpour2007}, magnetic semiconductors \cite{Ciccarelli2014}, and semiconductor quantum wells \cite{Ullrich2006,Wu2010}.

In this review we focus on systems where the spins are those of one species of itinerant carriers. For these systems, we explore the interplay between spin-coherence protecting mechanisms---such as chirality and, to a certain extent, spin-orbit coupling (SOC), see below---and sources of dissipation; we also emphasise the distinction between intrinsic and extrinsic mechanisms. The general framework will be that of two-dimensional electron liquids (2DEL).

We will distinguish intrinsic mechanisms of dissipation from extrinsic ones. We refer to purely electronic phenomena as `intrinsic', while the other mechanism are referred to as `extrinsic'. These may include deviations from a low-temperature perfect crystal or from a desired device, such as disorder, impurities, or interface roughness. We include dissipation due to phonons in the extrinsic category: in ideal samples extrinsic mechanisms can be suppressed, or at least their impact can be reduced by increasing the quality of the material or device or by reducing the temperature. Intrinsic mechanisms, by contrast, are introduced directly through the physical terms in the Hamiltonian needed to form the spin waves, spin plasmons, or the chiral spin waves. One of these mechanisms, called the Spin Coulomb Drag (SCD) \cite{Damico2000,Damico2010}, is due to the Coulomb interaction between the itinerant carriers.

\begin{table*}
\centering
\begin{tabular}{cccccc}
\hline\hline
    material & $m^*$ & $\epsilon _{r}$  & $a_{B}^{*}$ (nm) & $r_s$ & $\alpha k_{\rm F}/E_{\rm F} \times 10^{-2}$\\ \hline
    MoS$_2$ & $0.37$ & $6.5$ & $0.930$ & $19.19$ & -\\
    Si & $0.19$ & $7.7$ &  $2.14$ & $8.32$ & - \\
    CdTe & $0.105$ & $10.0$ &  $5.04$ & $3.54$ & $~0.4^*$\\
    GaAs & $0.067$ & $12.6$ &  $9.95$ & $1.79$ & $~0.2^*$ \\
    InSb & $0.02$ & $16.8$ &  $44.5$ & $0.401$ & $~3.7^*$\\
    LaAlO$_3$/SrTiO$_3$ & $3$ & $20000$ &  $353$ & $0.0506$ & $~1.6^\dagger$ \\
\hline\hline
\end{tabular}
\caption{Band mass $m^*$, dielectric constant $\epsilon_r$, Bohr radius $a_B^{*}$, Wigner-Seitz radius $r_s$ and spin-orbit strength $\alpha k_{\rm F}/E_{\rm F}$ for a set of typical 2D electronic materials. The 2D electronic density is taken as $n_{\mathrm{2D}}= 1.0\times 10^{11}{\rm cm}^{-2}$. The Rashba spin-orbit constant $\alpha$ has been evaluated from Ref.
\cite{Winkler2003} for (*) and from Ref. \cite{Caviglia2010} for ($\dagger$).}
\label{Tabrs}
\end{table*}

In recent years, condensed-matter physics has undergone a dramatic paradigm shift, triggered by the discovery of topological insulators
\cite{Kane2005,Koenig2007}. New and universal ways of characterizing band electrons through their topological properties have been
recognized as the key to understanding phenomena such as the spin Hall effect. In particular, chirality has emerged as a central protection mechanism for spin transport \cite{Hasan2010,Qi2010,Sinova2015} because it prevents backscattering. Chirality appears in inversion symmetry broken systems, such as electrons
confined in a quantum well plane subject to a perpendicular electric field \cite{Bercioux2015}. In such situations, the spin is locked to the electron momentum due to SOC. In addition to SOC, Coulomb many-body effects are then needed to form chiral spin waves \cite{Ashrafi2012}. But, as we will discuss, SOC can also become the cause of an intrinsic dissipation mechanism \cite{Dyakonov1986}.

2DELs are particularly well suited to explore the interplay between Coulomb interactions (direct and exchange), kinetic motion, and SOC and their related intrinsic dissipative mechanisms. They were first studied in Si inversion layers \cite{Ando1982}, then in high mobility III-V and II-VI quantum wells \cite{Davies1998,Harrison2005}, and more recently at oxide interfaces such as LaAlO$_3$/SrTiO$_3$ \cite{Stemmer2014} or in monolayers like Graphene \cite{Castro2009,Kotov2012} or from the dichalcogenide family like MoS$_2$ \cite{Wang2012,Butler2013}.

One can classify the systems listed above by the relative strengths of three protagonists: Coulomb interactions (both direct and exchange), kinetic energy, and SOC. The first important scaling parameter is the Wigner-Seitz radius $r_s$. It is defined as the ratio of the average electron-electron distance $\bar{d}$ to the electron Bohr radius $a_B^*=4\pi \epsilon_0 \hbar^2/(m^* {e^*}^2)$, where $m^*$ and $e^*$ are the material-dependent effective mass and screened effective charge, respectively. The Wigner-Seitz radius
estimates the ratio of the average Coulomb energy to the kinetic energy. Thus, high $r_s$ values correspond to Coulomb dominated systems with a strong collective behavior, while, on the other side, low $r_s$ values correspond to nearly noninteracting electrons. High $r_s$ can be reached by lowering the electron density, by increasing $m^*$, or by weakening the dielectric constant (which increases $e^*$). The relative strength of SOC to kinetic energy can be specified by the ratio $\alpha k_{\rm F}/E_{\rm F}$, where $\alpha$ is the typical Rashba constant (see Section \ref{sec:2.4}), $k_{\rm F}$ and $E_{\rm F}$ are the Fermi wavevector and energy. Table \ref{Tabrs} summarizes typical values of these parameters for various 2D systems for a given electron sheet density $n_{\mathrm{2D}}$.

\begin{table*}
\centering
\begin{tabular}{ll} \hline
abbreviation & meaning \\ \hline
2D & two-dimensional \\
3D & three-dimensional \\
2DEL & two-dimensional electron liquid \\
ALDA & adiabatic local-density approximation\\
CSR & chiral spin resonance\\
DFT & density-functional theory\\
DMI & Dzyaloshinskii-Moriya interaction \\
DP & D'yakonov-Perel'\\
EPR & electron paramagnetic resonance\\
ERRS & electron resonant Raman scattering\\
IRG & impulsive Raman generation\\
LDA & local-density approximation\\
LSDA & local spin-density approximation\\
SCD & spin Coulomb drag \\
SF-SPE & spin flip single-particle excitation\\
SFW & spin-flip wave\\
SOC & spin-orbit coupling \\
SP2DEL & spin-polarized two-dimensional electron liquid\\
TDDFT & time-dependent density-functional theory\\
TSG & transient spin grating\\
xc  & exchange-correlation\\
\hline
\end{tabular}
\caption{List of abbreviations.}
\label{Table_acro}
\end{table*}

The table shows how the same density of electrons can result in a system highly correlated by Coulomb interactions (MoS$_2$) or may correspond to nearly free particles (LaAlO$_3$/SrTiO$_3$). Si, CdTe or GaAs are intermediate. We will limit our discussion to these two last systems as they are well understood and very clean.
Prior studies referenced throughout this review have shown that the intrinsic mechanisms of dissipation discussed above are clearly visible in these systems.

This article is organised as follows: In Section 2 we set the stage by reviewing several relevant basic concepts such as exchange interactions in 2DELs,
the formation of various types of spin collective modes, SCD, and SOC in semiconductors; we also discuss the interplay between SOC and SCD, and summarize the
essential theoretical and experimental techniques to describe and probe the spin modes.
In Section 3 we  discuss collective spin modes in not-overall spin-polarized 2DELs that are influenced by Rashba and Dresselhaus
SOC. These modes can take place between two subbands (intersubband plasmons) or within one subband (chiral spin waves). In this section, we also introduce the excitation linewidth due to intrinsic dissipation and related formalism.
In Section 4 we then include effects of in-plane magnetic fields, considering 2DELs with a partially or fully polarized ground state and discussing spin-flip waves and their dispersions. As a special case,
we consider the spin-helix Larmor mode, which occurs in a 2DEL with equal-strength Rashba and Dresselhaus SOC, and
is an exact many-body result. Conclusions and some perspectives on future work are given in Section 5. Table \ref{Table_acro} presents a list of abbreviations used in this article.

\section{Important concepts and tools} \label{sec:2}
\subsection{Exchange in 2DELs}\label{sec:2.1}

In a 2DEL, Coulomb-exchange results from Pauli's exclusion principle and the Coulomb interaction between the electrons. The former prevents two electrons with parallel spin to be on top of each other. Thus, each electron is surrounded by a hole, the so-called ``Pauli-hole'' in the parallel-spin electron density. As electrons with parallel spin are repelled from each other, the Coulomb energy for parallel spins is reduced by an amount called the Coulomb-exchange, and this induces a self-alignment of spins. To first order, the ground-state Coulomb-exchange energy of massive electrons in a 2DEL (with parabolic dispersion) is universal and given by
\begin{equation}\label{Exch}
    \varepsilon_x=\frac{8\sqrt{2}}{6\pi r_s}R_y^*\left[\left(1+\zeta\right)^{3/2}+\left(1-\zeta\right)^{3/2}\right],
\end{equation}
where $R_y^*$ is the effective Rydberg energy and $\zeta$ is the spin-polarization degree of the 2DEL, $\zeta=\left(n_\ua-n_\da\right)/\left(n_\ua+n_\da\right)$. On the other hand, the ground-state kinetic energy reads
\begin{equation}
\varepsilon_K=\frac{1}{2r_s^2} R_y^*\left[\left(1+\zeta\right)^2+\left(1-\zeta\right)^2\right].
\end{equation}
We see immediatly that the ratio $\varepsilon_x/\varepsilon_K \propto r_s$.

It is important to note that one cannot map the Coulomb-exchange of itinerant electronic systems to the Heisenberg exchange constant that is encountered for localized orbitals. In a 2DEL, the motion of the electrons is a partner of the Pauli hole. However, when inserting magnetic impurities in a 2DEL, Heisenberg-type exchange occurs between the itinerant electrons of the 2DEL and the electrons localized on the sites of the magnetic impurities. For example, in CdMnTe doped quantum wells \cite{Gaj1979,Perez2009,PerezBook}, the Zeeman energy of conduction electrons has to be corrected by the Overhauser shift, which is the mean-field effect of this Heisenberg-type exchange. Thus, the full Zeeman energy can be written as
\begin{equation}
Z\left(  B_{\rm ext}\right)  =\Delta-\left\vert g^*\right\vert\mu_{\text{B}}B_{\rm ext} ,
\label{ZB0}%
\end{equation}
where $g^*$ is the effective g-factor, $\mu_{\rm B}$ is the Bohr magneton, and
\begin{equation}\label{Delta}
\Delta = J_{sd}\gamma
N_{\text{Mn}} | \langle {\hat{I}_{z}}\rangle\left(
B_{\rm ext},T\right) |.
\end{equation}
Here, $\gamma$ is the probability to find the electron in the quantum well, $J_{sd}$ is the $\mbox{$s$-$d$}$
exchange integral, $N_{\text{Mn}}$ is the
density of Mn spins, and $\langle {\hat{I}_{z}}\rangle\left(
B_{\rm ext},T\right) $ is the average spin of a single Mn atom at temperature $T$ and applied magnetic field $B_{\rm ext}$. Equation (\ref{ZB0}) underlines the
competition between the ``Overhauser shift'' $\Delta$ and the ``band'' Zeeman ($g^*$)
contribution, which appear with opposite signs in CdMnTe.

\begin{figure*}
\includegraphics[width=\linewidth]{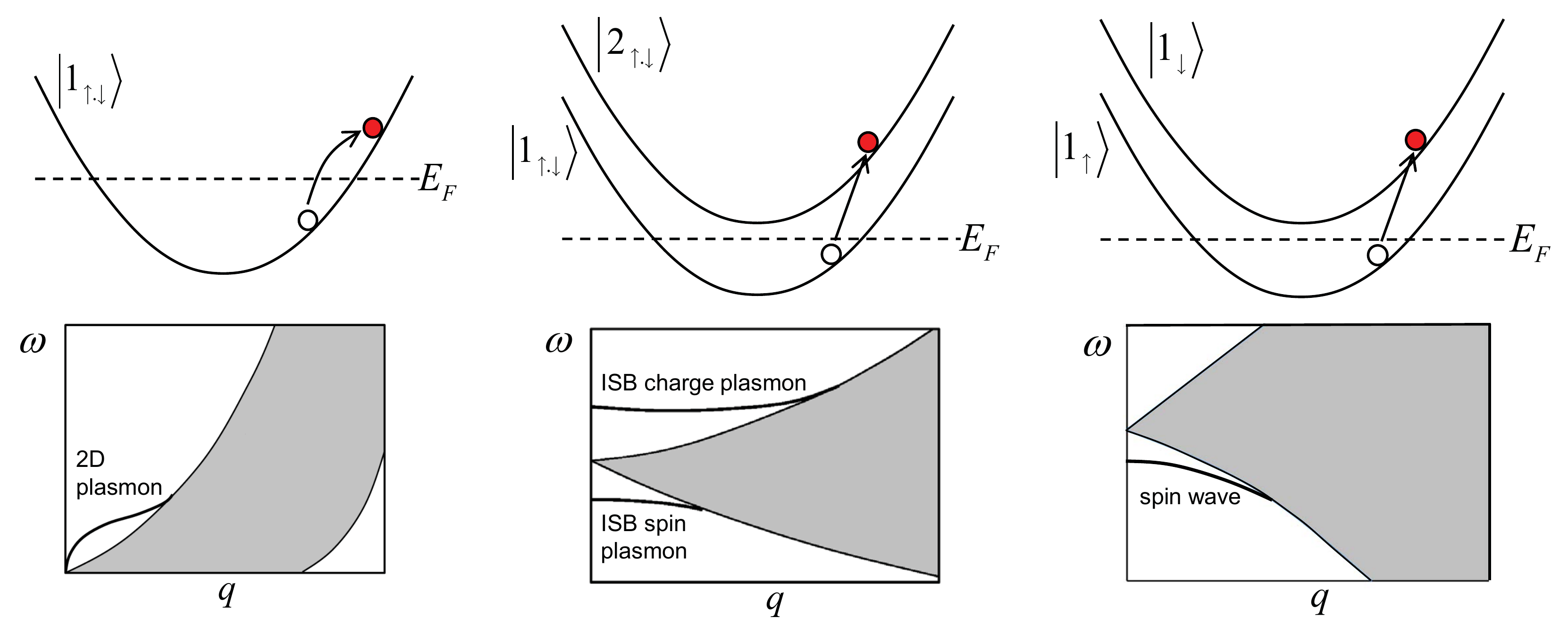}
\caption{ Left: intrasubband excitations in a non-spin-polarized 2DEL:
particle-hole continuum and 2D plasmon dispersion. Middle: intersubband excitations in a non-spin-polarized 2DEL:
particle-hole continuum and charge and spin plasmon dispersions.
Right: spin-flip excitations in a spin-polarized 2DEL: particle-hole continuum and spin wave.
SOC is not included.
} \label{fig1}
\end{figure*}

\subsection{Spin collective modes} \label{sec:2.2}

To discuss collective spin modes, it is conceptually helpful to begin with the excitations of a single electron in a two-level system,
$|1 \sigma\rangle \to |2\sigma'\rangle$, where $\sigma$ and $\sigma'$ are spin indices and 1,2 refer to orbital (subband) levels. Consider an inter-orbital state. Its time-dependent wave function is
\begin{equation}\label{wv}
\Psi(t) = \psi_1 \xi(\sigma) e^{-iE_1 t} + \lambda \psi_2 \xi(\sigma') e^{-iE_2 t} \:,
\end{equation} \label{2.2.1}
where $\lambda \ll 1$, $E_1$ and $E_2$ are the energies of the two levels, $\psi_{1,2}$ are the spatial parts of the wave functions (taken as real),
and $\xi$ are two-component spinors. We assume the $z$-axis as the direction of spin quantization.
For now, we ignore any effects due to SOC.

Let us calculate the first-order density and magnetization responses, $\delta n(t)$ and $\delta {\bf m}(t)$,
by substituting the wave function (\ref{wv}) into
$n = \mbox{tr}\{ \Psi \Psi^\dagger\}$ and ${\bf m} = \mbox{tr}\{ \bm{\sigma} \Psi \Psi^\dagger\}$,
where $\bm{\sigma}$ is the vector of Pauli matrices, and gathering contributions linear in $\lambda$.
If the excitation conserves spin, i.e., $\sigma=\sigma'$, then $\delta m_x(t)=\delta m_y(t)=0$ and
\begin{equation} \label{2.2.2}
\delta n(t)=\pm \delta m_z(t) =  2 \lambda \psi_1 \psi_2 \cos[(E_2-E_1)t].
\end{equation}
The charge-magnetization dynamics is {\em longitudinal}: it only involves components along the spin quantization axis.

For spin-flip excitations, i.e., $\sigma\ne \sigma'$, we find
$\delta n(t)=\delta m_z(t)=0$ and
\begin{eqnarray} \label{2.2.3}
\delta m_x(t) &=&  2 \lambda \psi_1 \psi_2 \cos[(E_2-E_1)t] \\
\delta m_y(t) &=&  \pm 2 \lambda \psi_1 \psi_2 \sin[(E_2-E_1)t] \label{2.2.4}
\end{eqnarray}
(the $+$ and $-$ signs in Eqs. (\ref{2.2.2}) and (\ref{2.2.4}) are for $\sigma'=\ua,\da$, respectively).
The magnetization dynamics is {\em transverse}, i.e., perpendicular to the spin quantization ($z$-)axis.

The basic findings of this simple example translate directly to the collective excitations in interacting many-electron systems
that are the subject of this review. One can distinguish several scenarios (to keep things simple, we do not include SOC here).
Figure \ref{fig1} gives an overview of the single-particle transitions and collective modes in a quasi-2DEL without and with applied magnetic field.
The general rule is that a collective mode will be stable and long-lived if its dispersion does not overlap with the particle-hole continuum,
since then, in the absence of disorder scattering, the mode cannot decay into single particle-hole pairs without violating energy-momentum conservation.
Decay into multiple particle-hole pairs is still possible, but much less effective.

Intrasubband transitions \cite{Ando1982}, see the left panel of Fig. \ref{fig1}, occur
within the same subband and take a particle from an occupied level (below the Fermi energy $E_{\rm F}$) to an unoccupied level outside the Fermi surface.
The associated collective mode is the intrasubband (or 2D) charge plasmon, whose dispersion $\omega(q)$ is shown in the bottom left panel of Fig. \ref{fig1}.
There is no corresponding intrasubband spin plasmon mode, where the spin-up and spin-down components of the 2DEL oscillate out of phase:
its dispersion lies entirely within the intrasubband particle-hole continuum and is therefore extremely short-lived \cite{Agarwal2014,Kreil2015}.

In an intersubband transition, see the middle panels of Fig. \ref{fig1}, the excitation occurs from an occupied subband level to an unoccupied level in a higher subband.
As shown, there is a charge plasmon whose dispersion lies above the intersubband particle-hole continuum,
and a spin plasmon below the continuum. The spin-conserving (longitudinal) and spin-flip (transverse) intersubband spin plasmons have the same frequency dispersions.

Intersubband charge and spin plasmons have been experimentally observed \cite{Pinczuk1988,Pinczuk1989,Gammon1990,Bao2004,Unuma2004,Unuma2006} and
theoretically investigated \cite{Ryan1991a,Ryan1991b,Chuang1992,Luo1993,Marmorkos1993} for three decades.

If the 2DEL is exposed to an in-plane magnetic field, then the spin-up and spin-down subbands are split. The corresponding intrasubband single-particle
transitions are shown in the right panel of Fig. \ref{fig1}. In contrast with the non-spin-polarized intrasubband plasmon case discussed above,
a collective long-lived spin-wave mode now exists.
The spin waves in a 2DEL have been experimentally and theoretically investigated \cite{Perez2009,Jusserand1992,Jusserand1993,Perez2007,Perez2011}.

The physical reason for the existence of the spin waves in the paramagnetic 2DEL is that the associated collective precessional motion of the electron spins in the
long-wave limit is protected by Larmor's theorem, as we will discuss in more detail below. For very low densities, the 2DEL undergoes a spontaneous
ferromagnetic phase transition \cite{GiulianiVignale,Attaccalite2002}; in that case, the spin waves (or magnons) become the Goldstone modes associated with the
spontaneous breaking of spin rotational symmetry.

\subsection{Spin Coulomb Drag} \label{sec:2.3}

In 2DELs, Coulomb interaction is at the origin of the collective modes introduced above. However, by giving rise to the SCD effect, it can also be a source of intrinsic dissipation for these modes. The SCD was proposed in 2000 \cite{Damico2000} and observed experimentally for the first time in a GaAs 2DEL in 2005 \cite{Weber2005}.

The spin-transresistivity \cite{Damico2000} couples two spin channels and is proportional, within the Kubo formalism, to the  response function between the corresponding spin current components \cite{Damico2000,Damico2013}. These may be spin-preserving `longitudinal' components (e.g. $\uparrow$ and $\downarrow$ spin current components) and/or spin-flipping `transverse' components (often referred to as `$+$' and `$-$' components, depending on their chirality). Accordingly, the Coulomb-originated contribution to the spin-transresistivity is divided into longitudinal SCD \cite{Damico2000, Damico2013} and transverse SCD \cite{Hankiewicz2008}. For the sake of simplicity, in what follows, we will use the acronym SCD to indicate longitudinal SCD and we will specify transverse or longitudinal only when necessary.

\begin{figure*}
\includegraphics[width=\linewidth]{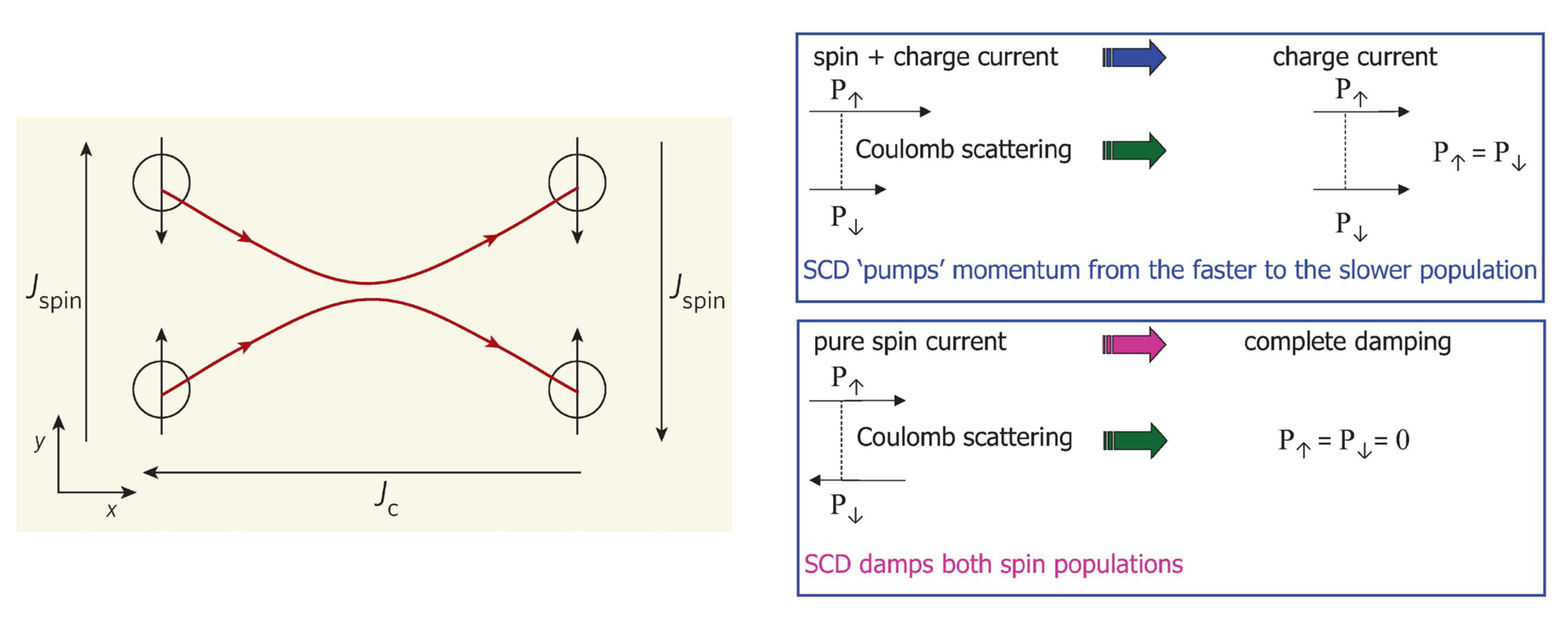}\centering
\caption{Left panel: schematic illustration of the  SCD mechanism for `head-on' scattering events.
\copyright 2005 Nature Publishing Group. Reprinted, with permission, from \cite{vanWees2005}.
Upper right panel: in a system with both spin and charge currents, the SCD will equilibrate the average momentum spin components leading to the persistence of the sole charge current.   Lower right panel: in the presence of pure spin currents and Coulomb interaction between the two spin populations, both average momentum spin components will be damped, eventually to zero, by the SCD. This is the situation in which the effect can be best measured experimentally.
All sketches in the left and right panels of the figure refer explicitly to the longitudinal SCD, where spin populations are characterized by $\uparrow$ and $\downarrow$ spins; however similar momentum transfer processes (and hence spin current decay) would apply to the case of transverse SCD, with  populations now defined by the `$+$' and `$-$' spin operators.}
\label{fig2}
\end{figure*}

The left panel of Fig. \ref{fig2} illustrates the microscopic mechanism of the SCD for the special case of a one-dimensional, head-on scattering event \cite{vanWees2005}:
due to Coulomb interaction, each electron in the pair experiences a conservation of its spin but a reversal of its momentum. This event does not alter the charge current (total momentum of the pair) but it reverses the spin current. Consider a spin-polarized 2DEL where a charge current travels together with a spin current: at the end, only the charge current will survive, as seen in the right-hand side of  Fig. \ref{fig2}.
The SCD is a many-body effect which stems from the non-conservation of the spin components of the total momentum in an electron liquid. Different spin populations will exchange momentum through Coulomb scattering leading, in the absence of a spin-dependent momentum ``pump'', to equal average momentum spin components, see Fig. \ref{fig2}.

Because of its Coulomb origin, the SCD translates into an intrinsic dissipation source for spin currents, which is the most effective for temperatures close to the Fermi temperature of the system, $T_{\rm F}=E_{\rm F}/k_B$, with $k_B$ the Boltzmann constant \cite{Damico2001, Damico2002}: for $T\ll T_{\rm F}$ the momentum space volume available for scattering decays quadratically, while at $T\gg T_{\rm F}$ the system behaves as noninteracting.
It follows that SCD will be negligible in metals, due to their high $T_{\rm F}$, but may become substantial in semiconductors, and especially for structures with lower dimensionality and $T_{\rm F}$ \cite{Flensberg2001,Damico2003}, where the spin-transresistivity $\rho_{\ua\da}$, which measures the strength of the effect \cite{Damico2000}, can become comparable or even higher than the Drude resistivity \cite{Weber2005,Yang2012}. While Coulomb interaction is key to the SCD, charge flow is not essential, so that the SCD will affect also pure spin currents (Fig.~\ref{fig2}, right lower panel).

In 2006 it was proposed that the SCD may contribute to the intrinsic linewidth of collective spin excitations \cite{Damico2006}, and the low-temperature frequency dependence of the spin transresistivity was analysed. Results showed that the SCD damping would be most effective for excitation energies comparable to $E_{\rm F}$. The intersubband longitudinal spin plasmon of a parabolic quantum well was proposed as a good candidate for observing the effect. This excitation in fact corresponds to an out-of-plane oscillation of the magnetization, with opposite spin components moving with opposite phases, as exemplified by Eq. (\ref{2.2.2}).

In 2007 and 2008 the transverse SCD was analysed and, together with the longitudinal SCD, was proposed as one of the mechanisms contributing to Gilbert damping in itinerant electron ferromagnets \cite{Hankiewicz2008,Hankiewicz2007}. Later the transverse SCD was explored as a source of intrinsic damping for transverse (spin-flip) spin waves \cite{Gomez2010} propagating in a high-mobility 2DEL. Here the SCD damping enters due to the coupling between the damping of the transverse spin current and the magnetization dynamics and it will be zero for $q\to 0$ (homogeneous limit).
The electron liquid was embedded in a Cd$_{1-x}$Mn$_x$Te/Cd$_{0.8}$Mn$_{0.2}$Te quantum well ($x<1\%$), where a highly polarized paramagnetic conductor is generated when a suitable magnetic field $B$ parallel to the quantum well surface is applied.  At finite $B$, this system supports a spin-flip wave between spin-split subbands, whose dispersion merges into the spin-flip single-particle excitation continuum at small values of the transferred momentum $q$ (right panel of Fig. \ref{fig1}). The system was analysed by Raman spectroscopy, which gives access to both dispersion relations. The lowest order inhomogeneous Gilbert damping contribution to the spin-flip wave linewidth is proportional to $q^2$ \cite{Hankiewicz2008,Gomez2010}, and includes contributions from both disorder and transverse SCD. While the $q^2$ damping rate behaviour was confirmed by the experiments \cite{Hankiewicz2008}, its transverse-SCD contribution was regarded to be too small to be relevant for the spin-flip wave lifetime.

At variance with the transverse SCD, the contribution of the longitudinal SCD to spin-plasmon damping remains finite in the homogeneous limit, and calculations based on the
three-dimensional local density approximation (3D-LDA) suggested \cite{Damico2006} that it should provide a sizable intrinsic contribution to the linewidth of intersubband spin plasmons, which should be measurable and dominant for clean quantum well samples.
Related experiments by inelastic light scattering were conducted a few years later, using electron liquids embedded in GaAs-based quantum well samples \cite{Baboux2012}. Results showed that 3D-LDA was providing an overestimate of the spin-plasmon linewidth. The SCD linewidth damping in spin plasmons is due to the decay of spin currents in the growth direction: the 3D-LDA overestimate demonstrated the necessity of a better treatment for both the 2D-3D crossover regime which occurs in quantum wells, as well as for inhomogeneous and non-local effects. These corrections to the theoretical approach will be discussed in Section \ref{sec:3.2}.

Further open experimental challenges with respect to the SCD will be discussed in Section \ref{sec:5}.

\subsection{Spin-orbit coupling} \label{sec:2.4}

Spin-orbit coupling  is a relativistic effect: electrons moving in a spatially varying electric field
experience a magnetic field in their own reference frame, which then interacts with the spin carried by the electrons \cite{BetheSalpeter}. Its expression in vacuum arises from the Pauli-Dirac equation:
\begin{equation}\label{HSOvac}
    \h_{\rm SO}^{\rm vac}=-\frac{e}{2m_0^2c^2}\hbS\cdot\bfE\times\hbp,
\end{equation}
where $m_0$ is the vacuum electron mass, $\bfE$ is the local electric field,  and $\hat {\bf S}$ is the electron spin operator.  SOC is naturally present in all materials, causing changes to the electronic structure, in particular for heavier elements and deep, strongly
bound levels, regardless of the crystalline symmetry of the system.

However, SOC can also have a strong influence on the itinerant carriers in valence and conduction bands,
which will be important for the collective spin modes that are of interest here. These SOC effects depend on
the crystal lattice structure: specifically, they require a breaking of inversion symmetry.
Recall the following important band-structure properties: time-reversal symmetry (which is preserved by SOC) leads to
$E_\ua(\bfk) = E_\da(-\bfk)$, and inversion symmetry causes $E_{\ua,\da}(\bfk) = E_{\ua,\da}(-\bfk)$ (here, $\bfk$ is the wavevector of the Bloch states).
Together, this gives rise to the spin degeneracy $E_\ua(\bfk) = E_\da(\bfk)$.

In the absence of inversion symmetry, it follows that the spin degeneracy of the bands is lifted.
A simple way of thinking about the resulting spin splitting is to
view it as a consequence of an additional term in the electronic Hamiltonian of the form
$\hat H_{\rm SO} = g^* \mu_B \bfB_{\rm SO}\cdot \hat {\bf S}$. Here,
$\bfB_{\rm SO}(\bfk)$ is an SOC-induced crystal magnetic field
which depends on the wavevector of the Bloch state it is acting on. Due to time-reversal symmetry we have $\bfB_{\rm SO}(-\bfk) = -\bfB_{\rm SO}(\bfk)$.

Inversion symmetry can be broken in several ways: by the crystal structure
itself, which is known as the Dresselhaus effect \cite{Dresselhaus1955}; through extrinsic electric fields which arise in structures such as
gated or asymmetrically doped quantum wells or inversion layers, which is known as the Rashba effect \cite{Bychkov1984,Bercioux2015,Manchon2015,Schapers};
and at interfaces with asymmetric bonds between non-common ions
\cite{Flatte2002}. The Dresselhaus and Rashba contributions tend to dominate for the systems considered here, so we will limit the discussion to these two effects.

\begin{figure*}
\includegraphics[width=\linewidth]{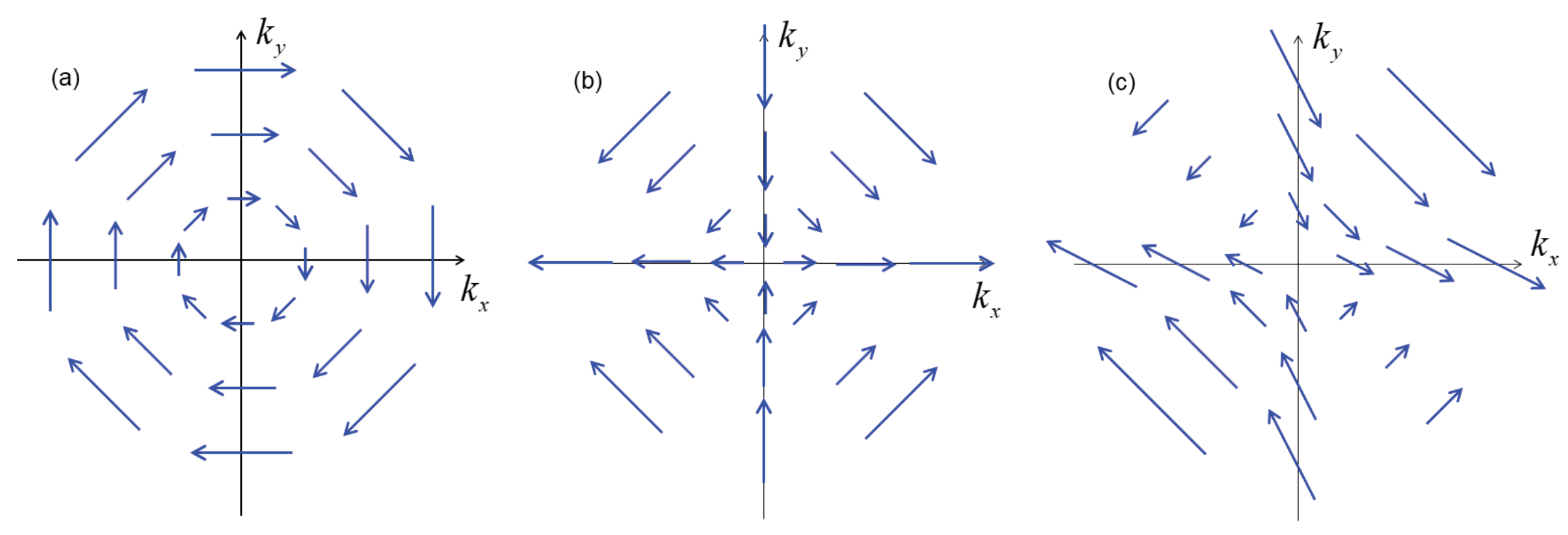}
\caption{ Spin-orbit effective magnetic fields in a 2DEL: (a) Rashba field, Eq. (\ref{Rashba}),  (b) Dresselhaus field, Eq. (\ref{Dresselhaus}),
(c) superposition of Rashba and Dresselhaus fields, with $\beta=2\alpha$.
} \label{fig3}
\end{figure*}

For typical III-V and II-VI semiconductors, the associated crystal magnetic fields for quasi-2D structures can be derived using standard perturbative techniques
known as $\bfk\cdot{\bf p}$ theory \cite{Winkler2003}. There is a dependence on the crystallographic direction of the 2D plane; we here limit ourselves
to zincblende quantum wells grown along the [001] direction (the corresponding expressions for other growth directions can be found in the review article
by Schliemann \cite{Schliemann2017}). One finds the following form for the Rashba crystal magnetic field:
\begin{equation} \label{Rashba}
\bfB_{\rm SO}^{\rm Rashba}(\bfk) = \frac{2 \alpha}{g^* \mu_B} \left(\begin{array}{c} k_y \\ -k_x \end{array}\right),
\end{equation}
while the Dresselhaus crystal magnetic field is
\begin{equation} \label{Dresselhaus}
\bfB_{\rm SO}^{\rm Dressel}(\bfk) = \frac{2 \beta}{g^* \mu_B} \left(\begin{array}{c} k_x \\ -k_y \end{array}\right) \:.
\end{equation}
Here, the 2D in-plane wavevector has the components $(k_x,k_y)$, where the $x$ and $y$ axes are aligned along the [100] and [010] directions, respectively. The Rashba and Dresselhaus coupling strengths, $\alpha$ and $\beta$, can be in principle be calculated via $\bfk\cdot{\bf p}$ theory
\cite{Eppenga1988,Andrada1994,Andrada1997,Pfeffer1997,Pfeffer1999a,Pfeffer1999b} (see Table \ref{Tabrs} for some examples) or using first-principles electronic structure methods \cite{Chantis2006}. The outcomes, however, are not always reliable, since many of the relevant system characteristics (such as well geometry or carrier concentration) are not very precisely known; it is often better to treat $\alpha$ and $\beta$ as fitting parameters.

The Rashba and Dresselhaus magnetic fields (\ref{Rashba}) and (\ref{Dresselhaus}) are schematically illustrated in Figs. \ref{fig3}a and b, respectively.
It can be seen that the Rashba field has a vortex-like structure, whereas the Dresselhaus field is anti-vortex-like.
The magnitudes of both fields, $B_{\rm SO}^{\rm Rashba}=2\alpha k/g^* \mu_B$ and $B_{\rm SO}^{\rm Dressel}=2\beta k/g^* \mu_B$, only depend on
$k = \sqrt{k_x^2 + k_y^2}$, not on the in-plane angle $\varphi = \tan^{-1}(k_y/k_x)$.

However, in systems where the Rashba and Dresselhaus effects are both present, the two crystal magnetic fields superimpose, as shown in
Fig. \ref{fig3}c for the case of $\beta=2\alpha$. In this case, the total field becomes dependent on the in-plane angle $\varphi$:
\begin{eqnarray} \label{9}
\lefteqn{
|\bfB_{\rm SO}^{\rm Rashba} + \bfB_{\rm SO}^{\rm Dressel}| }\nonumber\\
&=& \frac{2k}{|g^*| \mu_B}\sqrt{\alpha^2 + \beta^2 + 2\alpha\beta\sin 2\varphi}.
\end{eqnarray}
The dependence on $\sin 2\varphi $ will turn out to be very significant in our discussion of the spin modes.

SOC increases Coulomb scattering between different spin populations, and, as such, enhances dissipation due to SCD \cite{Tse2007}. SOC-enhanced longitudinal and transverse SCD is a source of intrinsic Gilbert damping \cite{Hankiewicz2007}, and may affect spin waves, including chiral spin waves \cite{Maiti2015a}
(see also Section  \ref{sec:3.3}).

This damping will persist at $q=0$ both in the presence or in the absence of an external magnetic field.
In fact the external magnetic field supports longitudinal spin-polarization, and hence longitudinal SCD, which is non-zero even at $q=0$ and would persist in the weak SOC limit.
However, even in the absence of a magnetic field, SOC couples the spin and orbital motion, and therefore, even at $q=0$ and in the absence of a magnetic field, spin waves in the presence of SOC are never pure spin excitations. Due to the coupling to orbital motion, momentum exchange between `$+$' and `$-$' spin populations is enhanced, leading to transverse SCD dissipation (see also Sections  \ref{sec:3.3} and \ref{sec:5}).

\subsection{D'yakonov-Perel' relaxation} \label{sec:2.5}

When a number of carriers (electrons or holes) with a distribution of wavevectors $\bfk$ are prepared in a given spin state in the host system
(metal or semiconductor), for instance as a ``spin packet'' via optical pumping \cite{Kikkawa1999} or via spin injection \cite{Johnson1985},
then the total spin of this nonequilibrium population of carriers will relax over time. Spin relaxation is an unavoidable phenomenon,
and plays an important practical role in spintronics \cite{Zutic2004,Schapers}.

Out of the various spin relaxation mechanisms that have been discussed in the literature \cite{Zutic2004},
we here focus on the D'yakonov-Perel' (DP) mechanism \cite{Dyakonov1986,Dyakonov1971}, since it raises an important point regarding
the nature of collective spin modes in semiconductors with SOC.

\begin{figure}
\includegraphics[width=\linewidth]{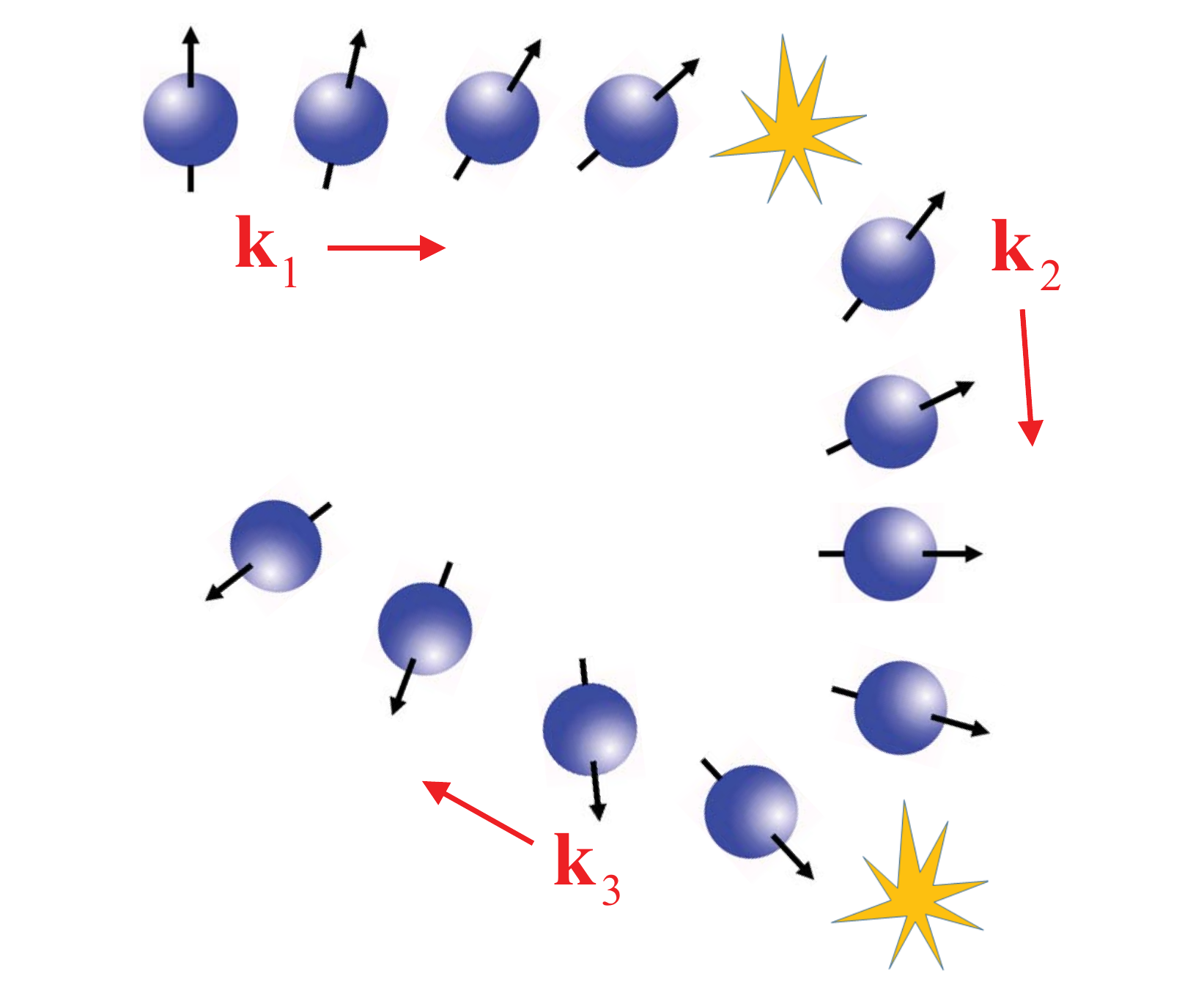}
\caption{ Schematic illustration of DP spin relaxation. Carrier spins precess about the spin-obit magnetic field $\bfB_{\rm SO}(\bfk)$
associated with their instantaneous wavevector $\bfk$. The precession changes after each scattering event.
} \label{fig4}
\end{figure}

DP spin relaxation occurs in materials where SOC causes the appearance of a wavector-dependent crystal magnetic field $\bfB_{\rm SO}(\bfk)$.
The spins of individual carriers  precess in the crystal magnetic field, but carriers with different $\bfk$ experience
a different $\bfB_{\rm SO}(\bfk)$, and hence precess at different rates and about different directions.
This leads to the dephasing of spin populations. The associated spin relaxation
time depends not just on $\bfB_{\rm SO}(\bfk)$, but also on momentum scattering; paradoxically, the shorter the momentum scattering time $\tau_p$
(related to collisions with impurities, phonons, and other electrons \cite{Mower2011,Marchetti2014a,Marchetti2014b}), the less
effective the DP mechanism is. The reason for this is that a higher rate of scattering events gives the carriers
less opportunity to precess between scattering events (this is called motional narrowing). The process is schematically illustrated in Fig. \ref{fig4}.

Thus, the DP mechanism causes a rapid dephasing of carrier spins---as long as the spins behave independently of one another, and simply
evolve in the $\bfk$-dependent SOC crystal magnetic field they find themselves in at a particular moment. As we will see below, this situation
changes dramatically if Coulomb many-body interaction effects are included.

\subsection{Theoretical techniques}\label{sec:2.6}

\subsubsection{Many-body Hamiltonian and single-particle states}\label{sec:2.6.1}

The 2DEL is a very widely studied model system, and a comprehensive account of
the theoretical techniques used to describe its properties would be beyond the scope of this review \cite{Ando1982,GiulianiVignale,Lipparini}.
Here, we just summarize the basic theoretical tools we need to describe collective spin modes in III-V and II-VI based, $n$-doped
quantum wells. In these systems, the electronic states are close to the bottom of the parabolic conduction band,
and are therefore well described within the effective-mass approximation. Thus,
we consider the many-body Hamiltonian
\begin{equation} \label{H}
\hat H = \sum_i^N \frac{ \hat {\bf p}_i^2}{2m^*} + \frac{1}{2}\sum_{i\ne j}^N\frac{e^{*2}}{|\bfr_i - \bfr_j|}
+ \hat H_{\rm SO} + \hat H_{\rm m}.
\end{equation}
Here, the first and second terms on the right-hand side are the kinetic and electron-electron interaction
Hamiltonians, respectively. The third term is the spin-orbit Hamiltonian,
\begin{equation}
\hat H_{\rm SO} = \mu_B \sum_i^N \bfB_{\rm SO}(\bfk_i) \cdot \bm{\hat\sigma}_i \:,
\end{equation}
where $\bm{\hat\sigma}_i$ is the vector of Pauli matrices associated with the spin of the $i$th electron,
and the SOC effective magnetic fields are due to the Rashba and Dresselhaus effects,
see Section \ref{sec:2.4}.

$\hat H_{\rm m}$ accounts for the influence of magnetic fields
on the itinerant conduction electrons. We only consider magnetic fields $\bfB_{\rm ext}$ that are in the plane of the 2DEL;
as long as the magnetic length $l_{\rm m}=\sqrt{\hbar/e B_{\rm ext}}$ exceeds the quantum well width,
the coupling of magnetic fields to the orbital motion (leading to Landau level quantization \cite{Li2011}) is suppressed and we only
need to include the Zeeman coupling of the magnetic field to the electron spins.

In addition to externally applied magnetic fields, $\hat H_{\rm m}$ can also account for the $\mbox{$s$-$d$}$ exchange coupling
between localized magnetic impurities and itinerant conduction electrons, see Section \ref{sec:2.1}. For simplicity,
we ignore these contributions in the present Section.

\begin{figure*}
\includegraphics[width=\linewidth]{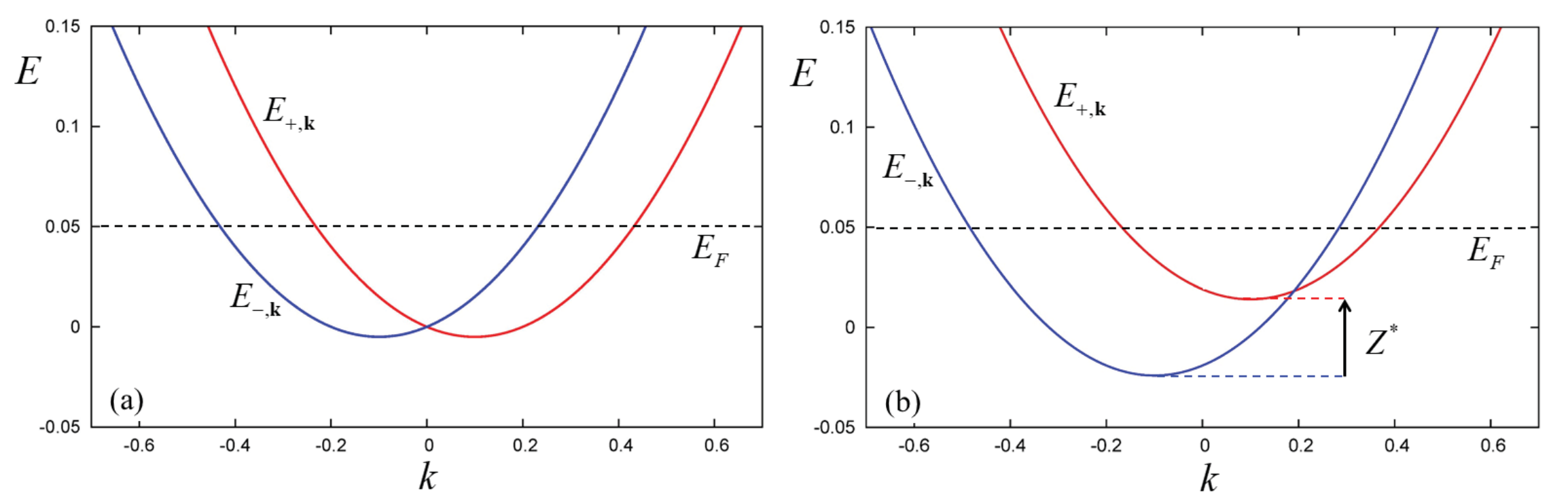}
\caption{ Single-particle energies for a 2DEL with Rashba and Dresselhaus SOC (assuming $\alpha=\beta=0.05$ and $\bfk$ along the [110] direction).
(a) No magnetic field. (b) Finite in-plane magnetic field ($Z^*=0.0381$).
} \label{fig5}
\end{figure*}

The electronic ground-state properties of the full many-body Hamiltonian (\ref{H}) can be obtained in various ways,
for instance using Landau Fermi liquid theory \cite{Ashrafi2013}.
A conceptually and computationally simpler alternative is density-functional theory (DFT)
\cite{Hohenberg1964,Kohn1965,Barth1972,Gunnarsson1976}. In DFT, the properties of the interacting electrons are calculated from a system of fictitious noninteracting Fermions moving in an effective self-consistent potential, including exchange-correlation (xc) contributions.
The single-particle wave functions $\Psi_{n\bfk}(\bfr)$ have a two-component spinor form:
\begin{equation}
\Psi_{n\bfk}(\bfr) = e^{i\bfk\cdot\bfr}\vec\psi_{n\bfk}(z) =
e^{i\bfk\cdot\bfr}\left( \begin{array}{c} \psi_{n\bfk\ua}(z)\\\psi_{n\bfk\da}(z)\end{array}\right),
\end{equation}
where $\bfk$ is a wavevector in the plane of the quantum well (assumed to be the $x-y$ plane), $n$ is a subband index,
and we include a $z$-dependence to allow for finite-width effects of the quantum well.
Using atomic units ($\hbar=m^*=e^*=1$),
the Kohn-Sham single-particle equation for the system (\ref{H}) is \cite{Karimi2017}
\begin{equation}\label{KS}
[\hat h_0 I +  \hat{\bf h}_{\rm SO,m}\cdot \bm{\hat\sigma}]\vec\psi_{n\bfk}(z) = E_{n\bfk}\vec\psi_{n\bfk}(z),
\end{equation}
where $I$ is the $2\times 2$ unit matrix,
\begin{equation}
\hat h_0 = \frac{k^2}{2} - \frac{1}{2}\frac{d^2}{dz^2} + v_{\rm conf}(z) + v_{\rm H}(z) + v_{\rm xc}^+(z)
\end{equation}
and
\begin{equation}
\hat{\bf h}_{\rm SO,m} = \mu_B \bfB_{\rm ext} + \mu_B\bfB_{\rm SO}(\bfk) + v_{\rm xc}^-(z) \hat e_{\bfB_{\rm ext}}.
\end{equation}
Here, $v_{\rm conf}(z)$ is the quantum well confining potential (e.g., a square well), $v_{\rm H}$ is the Hartree potential,
$v_{\rm xc}^\pm(z) = [v_{\rm xc \ua}(z) \pm v_{\rm xc \da}(z)]/2$ is the xc potential, and $\hat e_{\bfB_{\rm ext}}$ is a
unit vector along the in-plane magnetic field $\bfB_{\rm ext}$. The xc potential is approximated using the standard
local spin-density approximation (LSDA) \cite{Attaccalite2002,Perdew1992}.

The eigenstates of the Kohn-Sham equation (\ref{KS}) can be found analytically in the limiting case where the $z$-dependence can be
neglected \cite{Karimi2017}. The energy eigenvalues are
\begin{eqnarray} \label{Enk}
E_{\pm\bfk} &=& \frac{k^2}{2} +\frac{ \varepsilon_\ua+\varepsilon_\da}{2}
\pm k \bigg[\left(\frac{Z^*}{2k}+\beta\cos 2\varphi\right)^2
\nonumber\\
&+&
(\alpha+\beta \sin 2\varphi)^2 \bigg]^{1/2} ,
\end{eqnarray}
where $\varphi$ is the angle between $\bfk$ and the $x$-axis.
In Eq. (\ref{Enk}), $\varepsilon_\ua$ and $\varepsilon_\da$ are the spin-up and spin-down energy
eigenvalues of the Kohn-Sham system without SOC, and the renormalized (``dressed'') Zeeman energy is given by
\begin{equation}
Z^* = \varepsilon_\ua- \varepsilon_\da = Z + v_{\rm xc \ua} - v_{\rm xc \da} \:,
\end{equation}
where the ``bare'' Zeeman energy is $Z=g^* \mu_B B_{\rm ext}$, and we assume that the sign of the $\bfB_{\rm ext}$
is such that the $\ua$ states have higher energy than the $\da$ states.

Figure \ref{fig5} illustrates this for two different cases: (a) For $\bfB_{\rm ext}=0$ one obtains two parabolic bands
($+$ and $-$) which are horizontally displaced. (b) For $\bfB_{\rm ext}\ne 0$, the bands are also vertically displaced by $Z^*$.
In both cases, the states are filled up to the Fermi level $E_{\rm F} = \pi n_{\rm 2D} - (\alpha^2 + \beta^2)$.

In the absence of SOC, the two cases shown in Fig. \ref{fig5} reduce to the top left and top right energy bands
of Fig. \ref{fig1}, respectively.

\subsubsection{Calculation of spin-wave dispersions}\label{sec:2.6.2}

There are several theoretical methods for describing the collective spin dynamics in a 2DEL.
Going back to the work by Holstein and Primakoff \cite{Holstein1940}, one can define the spin-wave operator
(where $\hat\sigma^+ = \hat\sigma_x + i \hat\sigma_y$)
\begin{equation}\label{Sq}
\hat S_\bfq^+ =\frac{1}{2}\sum_i \hat\sigma_i^+ e^{i\bfq\cdot \bfr_i} \:,
\end{equation}
whose equation of motion is given by
\begin{equation}\label{Sqmot}
i\frac{d}{dt}\hat S_\bfq^+ = [\hat S_\bfq^+,\hat H] \:.
\end{equation}
This formal relation provides the starting point for a full account of the interplay between electronic many-body effects,
SOC and magnetic-field effects in the spin-wave dynamics \cite{Perez2009,Perez2011,Karimi2017,Perez2016,Karimi2018}.

A connection to linear-response theory can be made by defining the transverse (or spin-flip) response function \cite{GiulianiVignale,Perez2011}
\begin{equation}\label{chi_trans}
\chi_{\da\ua,\da\ua}(\bfq,\omega) = \langle\langle S_\bfq^+;S_\bfq^-\rangle\rangle_\omega \:,
\end{equation}
where $\omega$ is the frequency,
and $\langle\langle \ldots ; \ldots \rangle\rangle_\omega$ denotes a frequency-dependent response function defined in the standard way \cite{GiulianiVignale}.
Here, we consider the time-dependent spin-density matrix
\begin{equation}\label{2.7}
n_{\sigma\sigma'}(\bfr,t) = \langle \Psi(t) | \hat \psi_{\sigma'}^\dagger(\bfr)\hat \psi_\sigma(\bfr) | \Psi(t)\rangle
\end{equation}
as basic variable, where $\Psi(t)$ is the full many-body wave function [associated with the Hamiltonian $\hat H$, Eq. (\ref{H}), plus a
perturbation], and $\hat \psi_{\sigma}(\bfr)$ and $\hat \psi_{\sigma}^\dagger(\bfr)$ are Fermionic field operators for spin $\sigma$.

Within time-dependent density-functional theory (TDDFT) \cite{Ullrich2012}, the linear response of the spin-density matrix is given by
\begin{equation}\label{2.9}
n_{\sigma\sigma'}^{(1)}(\bfq,\omega) = \sum_{\tau\tau'} \chi_{\sigma\sigma',\tau \tau'}(\bfq,\omega)
v_{\tau\tau'}^{(1)\rm eff}(\bfq,\omega) \:,
\end{equation}
where $\chi_{\sigma\sigma',\tau \tau'}(\bfq,\omega)$ is the response function of the corresponding noninteracting 2DEL,
and the effective perturbation is
\begin{eqnarray}\label{2.10}
\lefteqn{\delta v_{\tau\tau'}^{(1)\rm eff}(\bfq,\omega)
=
v_{\tau\tau'}^{(1)}(\bfq,\omega)} \\
&+&
\sum_{\lambda\lambda'} \left[ \frac{2\pi}{q} + f^{\rm xc}_{\tau\tau',\lambda\lambda'}(\bfq,\omega)\right]
n_{\lambda\lambda'}^{(1)}(\bfq,\omega). \nonumber
\end{eqnarray}
Here, $f^{\rm xc}_{\tau\tau',\lambda\lambda'}(\bfq,\omega)$ is the xc kernel for the spin-density matrix
response of the 2DEL, which can be calculated using the LSDA \cite{Ullrich2002,Ullrich2003}, in which
case it becomes independent of $\bfq$ and $\omega$.

To obtain the full excitation spectrum of the electronic system, one sets the external perturbation $v_{\tau\tau'}^{(1)}(\bfq,\omega)$ to zero,
so that only the self-consistent Hartree and xc perturbations remain in Eq. (\ref{2.10}). Solving the response equation (\ref{2.9})
then yields the single-particle excitations and collective modes. By expanding the noninteracting response function $\chi_{\sigma\sigma',\tau \tau'}(\bfq,\omega)$
in orders of $\bfq$ one can obtain analytic results for the mode dispersions. We will come back to this later, in Sections \ref{sec:4.3} and \ref{sec:4.4}.

Instead of TDDFT, it is also possible to calculate the transverse response function (\ref{chi_trans}), and the spin-wave properties following from it
(including dissipation), using
Fermi-liquid theory \cite{Brataas1997} or diagrammatic many-body techniques \cite{Maiti2015a,Mishchenko2004,Maiti2015b,Maiti2016,Maiti2017}.

Furthermore, more phenomenological descriptions of the collective spin dynamics in a 2DEL can be obtained via Landau Fermi liquid theory \cite{Ashrafi2012,Iqbal2014}
or via the Landau-Lifshitz-Gilbert equations of motion \cite{Lifshitz}.

\subsection{Experimental techniques}\label{sec:2.7}

Probing the spin degrees of freedom of a 2DEL can be done directly and similarly to the nuclear magnetic resonance, where a microwave cavity-mode magnetic field $\bfb(\bfr,t)$ oscillates with a frequency $\omega$ in the plane perpendicular to the polarizing magnetic field $\bfB_0$. The typical perturbing Hamiltonian reads: $\hat h_{\rm d}=-g^\ast\mu_B\hat S_{-\bfq}^-b(t)_{\bfq}^+$, where $\hat S_{-\bfq}^-$ is the spin-wave operator introduced in (\ref{Sq}), and $b(t)_{\bfq}^+=\int (b_x(\bfr,t)+ib_y(\bfr,t))e^{i\bfq\cdot\bfr}d\bfr$ is the spatial-Fourier transform of the transverse oscillating magnetic field. In general, the typical variation length scale of $\bfb(\bfr,t)$ is much larger than the electron wavelength; thus, this technique, called the electron paramagnetic resonance (EPR), probes only the macroscopic spin motion of $S_{\bfq=0}^+$. Because the frequencies of the magnetic field match the discrete cavity modes, the EPR response is given by the absorption spectra obtained by sweeping the amplitude and/or direction of the static polarizing magnetic field $\bfB_0$. It is proportional to the imaginary part of the transverse spin susceptibility $\Im\chi_{\da\ua,\da\ua}(\bfq=0,\omega)$ defined in Eq. (\ref{chi_trans}).

The typical detection threshold of a standard EPR setup is around $10^{11}$ spins placed in the cavity. In the absence of SOC, the macroscopic spin oscillates at the Larmor frequency $\omega_0=g^*\mu_B B_0$ which, remarkably, is independent of electron-electron interactions: this is the Larmor theorem \cite{Perez2007,Jusserand2003} (see below).
In such case, the outputs of EPR measurements are the determination of the band g-factor $g^*$ and its anisotropy as in Ref. \cite{Giorgioni2016}. The linewidth of the resonance at $\omega=\omega_0$ is also related to the homogenous-mode relaxation rate $1/T_2$, which is defined from the effective (Bloch) equation of motion that can be inferred from Eq. (\ref{Sqmot}),
\begin{equation}
i\frac{d}{dt}\hat S_{\bfq=0}^+ = \omega_0\hat S_{\bfq=0}^+-i\hat S_{\bfq=0}^+/T_2 \:.
\end{equation}

We discuss in Section \ref{sec:4.4} the Larmor theorem when the spin-rotational symmetry is broken by SOC.

\begin{figure*}
\includegraphics[width=\linewidth]{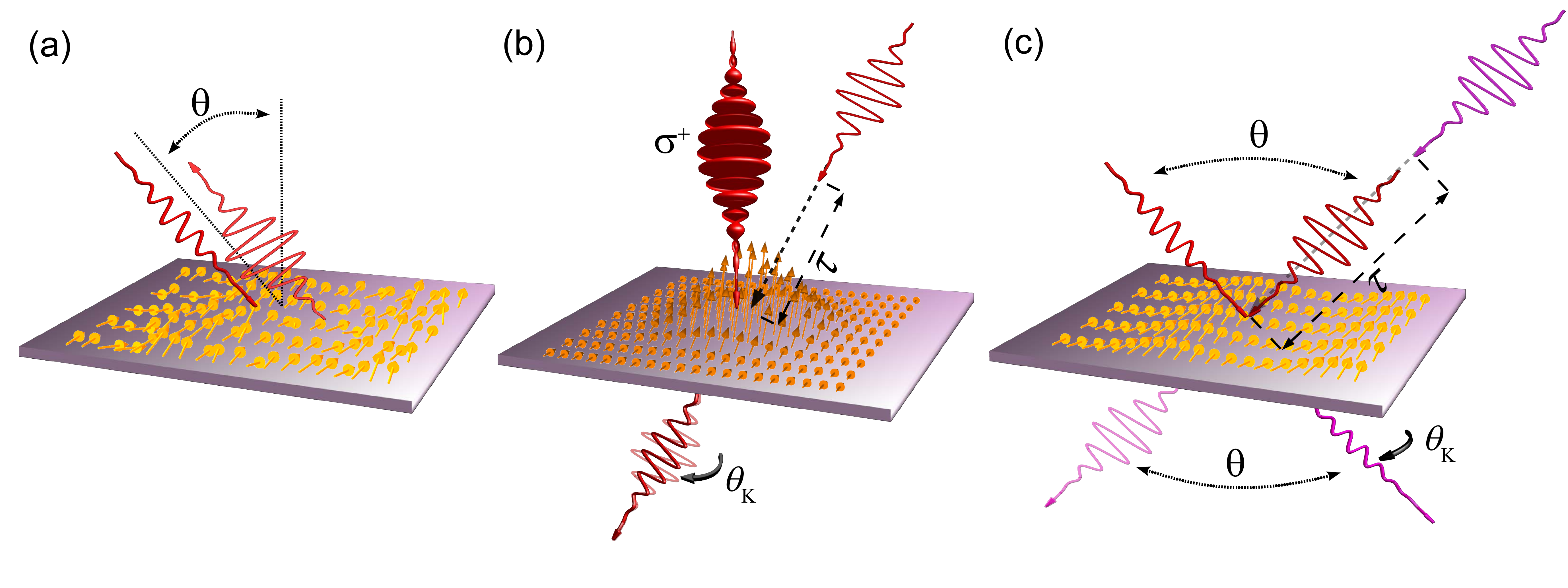}
\caption{(a) ERRS experiment: incoming linearly polarized photons hit the 2DEL in its thermal equilibrium. Fluctuations of the spins lead to backscattered photons with crossed polarization. Selecting the angle of incidence and backscattering (here both equal to $\theta$) probes the fluctuation spectrum at momentum $q\simeq\frac{4\pi}{\lambda}\sin\theta$, where $\lambda$ is the photon wavelength. (b) IRG experiment: a circularly polarized laser pulse creates a coherent spin state in the 2DEL. After the pulse, spins are out of equilibrium and the spin state evolves freely in time. A $\tau$-delayed linearly polarized pulse experiences Kerr rotation through transmission (or reflection). The Kerr angle $\theta_K$ is proportional to the out-of-plane spin component. (c) TSG experiment: here, the circular pulse of IRG is divided into two crossed linearly polarized pulses, with different angles of incidence (here $\pm\theta/2$). They generate a coherent spin state, but, opposite to IRG, an in-plane momentum $q\simeq\frac{2\pi}{\lambda}\sin\theta$ is transferred to the spin excitation, which generates a spin grating. The delayed probe is diffracted by the spin grating while experiencing rotation of the polarization (here $\theta_K=\pi/2$).} \label{fig6}
\end{figure*}

Electromagnetic waves can indirectly couple to the spin modes through the $\hat h_{\rm ind}=-(e/m^*)\bfA\cdot\hbp$ coupling. Here, $\bfA(\bfr,t)$ is the electromagnetic vector potential and $\hbp$ is the electron momentum. Despite the fact that the spin degrees of freedom do not appear in $\hat h_{\rm ind}$, an indirect coupling arises from the spin-orbit coupling $\hbL\cdot\hbS$ in the host crystal, which creates split-off bands with spin-mixed states. For example, consider the total momentum $\hbJ=\hbL+\hbS$ in crystals for $p$-bands ($l=1,J=3/2,1/2$); the $J=1/2$ states are of mixed spin and read
\begin{eqnarray}
    \left\vert J=\frac{1}{2},J_{z}\pm\frac{1}{2}\right\rangle &=&-\sqrt{\frac{2}{3}}\left\vert p_z,S_{z}=\pm\frac{1}{2}\right\rangle\\
&\pm&\sqrt{\frac{1}{6}}\left\vert \left(  p_x\pm ip_y\right),S_{z}=\mp\frac{1}{2}\right\rangle. \nonumber
\end{eqnarray}
If the 2DEL occupies $s$-bands of the same crystal, consider the process described by the optical matrix element $\langle l=0,S_{z} | \bfA\cdot\hbp | J=1/2, J_{z}=1/2\rangle$: an electromagnetic vector potential polarized along the $z$ axis couples to an $| l=0,S_{z}=+\frac{1}{2}\rangle$ electron in the 2DEL, while an $x$-polarized one couples to a spin-down electron \cite{Kane1957}. Hence, coupling to transverse spin modes in the 2DEL can be achieved via
second-order spin-flip processes characterized by matrix elements such as
\begin{eqnarray}\label{2nd}
 M_{\uparrow \downarrow} &=& \left\langle l=0,S_{z}=+\frac{1}{2}| A_z\hp_z | \frac{1}{2},\frac{1}{2}\right\rangle \nonumber\\
 &\times& \left\langle \frac{1}{2},\frac{1}{2}| A_x\hp_x | l=0,S_{z}=-\frac{1}{2}\right\rangle.
\end{eqnarray}
Experimental techniques involving this process are Electronic Resonant Raman Scattering (ERRS) \cite{Jusserand1992,Perez2007,Pinczuk1992}, Impulsive Raman generation (IRG) \cite{Bao2004,Rungsawang2013}, and Transient Spin-Gratings (TSG) \cite{Weber2005,Koralek2009}. They are sketched in Fig. \ref{fig6}.

ERRS is a continuous-wave optical spectroscopy technique where energy and momentum are conserved. Incoming photons from a monochromatic optical beam (laser) scatter with electrons in the crystal and the spectrum of scattered photons is measured with a spectrometer. The matrix element (\ref{2nd}) yields the scattering probability (i.e, the
Raman cross section). In the ERRS case, only one of the vector potentials  in Eq. (\ref{2nd}) belongs to the incoming laser beam, the second one should be viewed as the vacuum electromagnetic field because this matrix element describes a spontaneous process.

In general, the spectrum shows ``Raman lines'' at an energy below the Rayleigh line. The latter is due to elastic scattering of the incoming photons. The Raman shift is the energy difference between the Rayleigh line and the Raman line. It corresponds to an excitation energy of the crystal, which is then measured from this Raman shift. Raman lines are discriminated from other photonic lines by tuning the laser wavelength (if possible): the Raman lines follow the Rayleigh lines by a constant shift.

Discrimination between electronic and other processes (vibrational) underlying the presence of a Raman line is sometimes tricky. In general, for 2DELs, electronic Raman lines are broader (because electronic excitations live shorter than phonons), disappear quickly when raising the temperature of the system from $1.0 {\rm K}$ to $20 {\rm K}$ and, furthermore, they are strongly resonant. This means that the electronic Raman line is visible only when the incoming photon wavelength is close to an optical resonance of the crystal.

Identification of the involved excitation (plasmon, spin-plasmon, spin-wave etc...) is done by (i) analysing the selection rules followed by the polarizations of the incoming and scattered photons (as developed above) and  (ii) measuring the dependence of the Raman line with the transferred momentum (if possible by the experimental setup and the dimensionality of the electronic system). Comparison with theory is the last necessary step to complete the identification. In practice, electronic Ramanists construct their reasoning with the non-resonant approximation: when neither the incoming nor the scattered photons are in resonance with any of the electronic transitions in the crystal,  the ERRS cross section is proportional \cite{Pinczuk1988} to $\Im\chi_{\da\ua,\da\ua}(\bfq,\omega)$, which can be calculated within the frame developed in Section~\ref{sec:2.6}. In the particular case of a 2DEL, ERRS allows the measurements of the dispersions of spin modes \cite{Baboux2012,Perez2016,AkuLeh2011,Baboux2013} by varying the angle of incidence $\theta$ of the incoming and scattered photon with respect to the 2DEL plane\footnote{In the Raman process, the momentum is conserved. In 2D and 1D, one can probe excitations with a well defined momentum $\bfq$ by varying the momenta of the incoming $\bfk_i$ and scattered $\bfk_s$ photons, as $\bfq=(\bfk_i-\bfk_s)_{\rm conserved}$, where the subscript refers to the components of the momentum that are conserved by the dimensionality. In 2D, in a backscattering geometry $\bfk_i\simeq-\bfk_s$, $q\simeq\frac{4\pi}{\lambda}\sin\theta$, where $\theta$ is the incidence angle of the photons with respect to the 2D plane.}.

IRG is a transient time-domain spectroscopy. A circularly polarized laser pulse of duration $\tau$ hits the crystal at time zero. It leaves the 2DEL in a state of the form $|t=\tau\rangle=|0\rangle + c_f|\rm sw\rangle$, which is a coherent state between the ground state and a spin state (spin-plasmon, spin-wave). The amplitude $c_f$ is given by the matrix of the second-order process (\ref{2nd}). Here, the two vector potentials belong to the same beam. The circular polarization provides the two required cross-polarized photons involved in the matrix element (\ref{2nd}). Thus, during the pulse, the Raman process acts as a coupling between opposite spin states of the 2DEL.  As the two involved photons have necessarily the same momentum, with same incidence and direction, no momentum is transferred during the process, and only $\bfq=0$ spin modes can be excited. After that, the coherent state evolves freely. The expectation values of the transverse spin components $\langle t=\tau|\hbS_{\bfq=0}^+|t=\tau\rangle$ oscillate at the frequency $\omega_0$ and decay within a time $T_2$.   The oscillation of the transverse spin components can be probed by measuring the rotation of the polarization of a linearly polarized delayed laser pulse which crosses the sample \cite{Bao2004}. The latter effect is referred as the magneto-optical Kerr effect (MOKE).

TSG involves two crossed linearly polarized laser pulses. Contrary to IRG, the two laser beams hit simultaneously the 2DEL plane with different angles of incidence. Thus, an in-plane momentum can be transferred to the induced spin excitation, which is called a transient spin-grating. The mechanism is still described by the matrix element (\ref{2nd}): here, the two vector potentials belong to each of the two beams. In this case, spin components at non-zero $\bfq$ will oscillate and decay. The expectation values $\langle t=\tau|\hbS_{\bfq}^+|t=\tau\rangle$ are the spin-grating. Similarly to IRG, the dynamics of these components will be sampled by a linearly polarized, delayed pulse. Photons of this pulse have a momentum $\bfk_p$ and can be diffracted by the transient spin grating, such that the diffracted beam has in-plane momentum $(\bfk_p)_{\rm conserved}\pm\bfq$. At the same time, due to the Kerr effect, the diffracted and the probe beams are cross-polarized \cite{Weber2005,Koralek2009}.

A comparison of the efficiency between direct (EPR) and indirect (ERRS, IRG, TSG) coupling to spin degrees of freedom can be done roughly by considering the ratio $\left\vert\hat h_{\rm d}/\hat h_{\rm ind}\right\vert\simeq (\alpha^*)^{-1}$, where $\alpha^*$ is the material fine structure constant. The inverse $(\alpha^*)^{-1}$ appears after converting the $\bfA\cdot\hbp$ coupling into the $\bfE\cdot\hbr$ coupling and assuming $\langle\bfr\rangle\simeq a_B^{e-h}$, where $\bfE$ is the electric field of the electromagnetic wave, $\bfr$ is the electron position operator and $a_B^{e-h}$ is the typical Bohr radius of an electron-hole pair in the crystal. Thus, the optical resonance makes the indirect coupling more sensitive.

\section{Spin-unpolarized 2DEL}\label{sec:3}

\subsection{Intersubband spin plasmons: collective spin-orbit effects} \label{sec:3.1}

\begin{figure*}
\includegraphics[width=\linewidth]{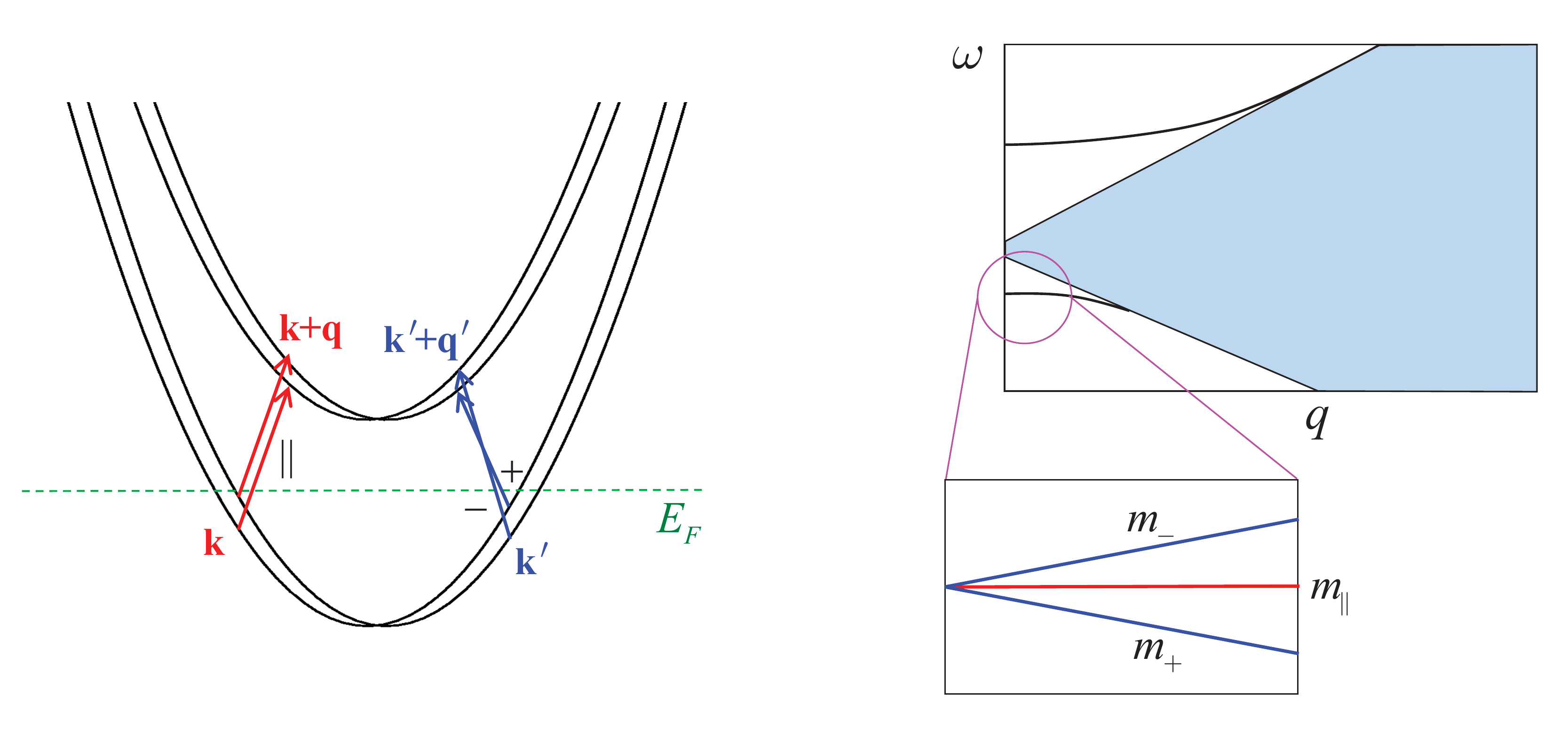}
\caption{Left: schematic representation of the two lowest SOC-split subbands of a quantum well, with longitudinal (red arrows)
and transverse (blue arrows) single-particle transitions.
Right: Associated intersubband particle-hole continuum and charge and spin-plasmon dispersions.
The closeup reveals a three-fold splitting of the intersubband spin-plasmon dispersion into one longitudinal and two transverse collective modes.
} \label{fig7}
\end{figure*}

As we discussed in Section \ref{sec:2.2} (see Fig. \ref{fig1}), the non-spin-polarized 2DEL can sustain intersubband spin plasmon modes.
We now ask how these modes are influenced by the presence of SOC \cite{Baboux2012,Ullrich2002,Ullrich2003,Ullrich2013}.
A collective spin mode is a coherent superposition of many single-particle
spin excitations of the 2DEL; the left-hand side of Fig. \ref{fig7} shows four different excitations between the two lowest, SOC-split subbands,
assuming for simplicity that both subbands have the same parabolicity, and are subject to the same $\bfB_{\rm SO}(\bfk)$.
We distinguish longitudinal ($||$) and transverse ($+,-$) single-particle excitations (shown as red and blue arrows, respectively):
the $||$  excitations are $E^{(1)}_{\pm \bfk}\to E^{(2)}_{\pm (\bfk+\bfq)}$,
and the $\pm$  excitations are $E^{(1)}_{\pm \bfk'}\to E^{(2)}_{\mp( \bfk'+\bfq')}$.

Intersubband single-particle excitations with different $\bfk$ but the same momentum transfer $\bfq$ all have a different energies, which gives rise to the intersubband particle-hole continuum, shown as shaded area on the right-hand side of Fig. \ref{fig7}. The intersubband charge and spin plasmons, on the other hand,
are collective modes which
are ``held together'' by Coulomb interactions.

Based on the discussion in Section \ref{sec:2.5}, one would expect the DP mechanism to play an adverse role
for the intersubband spin plasmons: the underlying single-particle spin excitations are each subject to different $\bfB_{\rm SO}$, which
should lead to a significant line broadening due to precessional dephasing. However, this is not the case: in the absence
of impurities, defects, phonons, and dissipative electron-electron interactions (such as the SCD, see Section \ref{sec:3.2.2}),
the intersubband spin plasmons are sharp lines, since the presence of Coulomb many-body interactions renders the precessional dephasing ineffective.

\begin{figure}
\includegraphics[width=\linewidth]{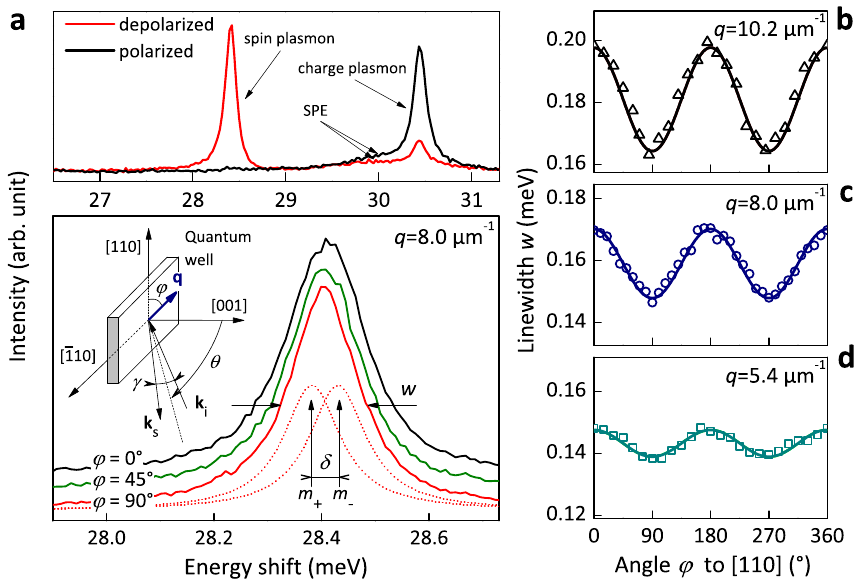}
\caption{Anisotropic splitting of intersubband spin plasmons in a GaAs quantum well. (a) Top: inelastic light scattering spectra of
charge and spin plasmons and single-particle excitations (SPE). Bottom: the quasi-Lorentzian peak of the intersubband spin plasmon is
the sum of the $m_+$ and $m_-$ transverse modes split by an amount $\delta$. (b-d) Variation of the linewidth $w$ with
in-plane angle $\varphi$ for three values of $q$. The experimental data points are reproduced by TDDFT linear response theory.
\copyright 2012 American Physical Society. Reprinted, with permission, from \cite{Baboux2012}.
} \label{fig8}
\end{figure}

The experimental proof of this remarkable phenomenon was given in Ref. \cite{Baboux2012}, see Fig. \ref{fig8}.
Inelastic light scattering (ERRS) reveals rather sharp intersubband charge and spin plasmon peaks. The width of the spin plasmon peak
increases linearly with the magnitude of the plasmon wavevector, $q$, and exhibits a modulation as a function of the in-plane direction of $\bfq$.
This behavior of the intersubband spin plasmon linewidth is a direct consequence of SOC.

According to theoretical predictions \cite{Ullrich2002,Ullrich2003}, the intersubband spin plasmon dispersion is split into three branches,
as illustrated on the right-hand side of Fig. \ref{fig7}.
The longitudinal intersubband spin plasmon dispersion $\omega_{||}(\bfq)$ is independent of SOC to within the lowest order of the Rashba and Dresselhaus
coupling constants $\alpha$ and $\beta$. The two transverse intersubband spin plasmon dispersions, on the other hand, are given by
\begin{eqnarray}
\omega_\pm(\bfq) &=& \omega_{||}(\bfq) \pm q C\sqrt{\alpha^2 + \beta^2 + 2\alpha\beta \sin2\varphi_q} \nonumber\\
&+& {\cal O}((\alpha,\beta)^2) \:,
\end{eqnarray}
where $\varphi_q$ is the angle between $\bfq$ and the $[100]$ direction, and
$C$ is a constant that depends on the subband envelope functions, the density of electrons, and on the xc kernel $f^{\rm xc}$.
The experimentally measured intersubband spin plasmon peak is a composite of the two intersubband spin-flip plasmons, $m_+$ and $m_-$.
Thus, the intersubband spin plasmon splitting is, to lowest order in $q$, $\alpha$ and $\beta$,
\begin{equation}
\delta(\bfq) = 2q C\sqrt{\alpha^2 + \beta^2 + 2\alpha\beta \sin2\varphi_q} \:.
\end{equation}
Clearly, $\delta$ grows linearly with $q$ and has an amplitude that is modulated with period $\pi$, in agreement with
the experimental findings. We briefly mention that to second and higher order in SOC, one finds additional contributions to
the splitting between the longitudinal and transverse modes which remain nonvanishing even at $\bfq=0$ but are very small \cite{Ullrich2003}.

The physical picture that emerges from these observations is thus as follows: the intersubband spin plasmon behaves like a
macroscopic magnetic moment which precesses in a collective SO magnetic field $B_{\rm SO}^{\rm coll}(\bfq)$, and whose magnitude
is enhanced by a factor $C$ compared to the bare Rashba and Dresselhaus SO magnetic field $B_{\rm SO}(\bfq)$, defined in Eq. (\ref{9}).
The intersubband spin plasmon fine structure can thus be viewed as an intrinsic normal Zeeman effect \cite{Ullrich2013}:
the three-fold splitting of the plasmon dispersion finds a direct analogy to the so-called ``Lorentz triplet''
of atomic spectroscopy, where spectral lines are split according to the selection rules $\Delta m_s=0$ and $\Delta m_l = 0,\pm 1$.
This picture was further experimentally confirmed by mapping out $B_{\rm SO}^{\rm coll}(\bfq)$ using an applied external in-plane magnetic field
\cite{Baboux2012}, which showed that the many-body enhancement of the collective SO magnetic field over the bare SO field is about a factor of five.

\subsection{Dimensionality crossover and nonlocality} \label{sec:3.2}
There are challenges in calculating dissipation and excitation linewidths in many-body systems from first principles. For example, the most widely used (time-dependent) DFT approximation for calculating excitation spectra is adiabatic-LDA (ALDA), which describes a Markovian dynamics local in time and space and hence does not account for dissipative effects nor for strong inhomogeneities.  However the nanoscale and low-dimensional systems typically proposed for spintronics and quantum technology applications, may display strong inhomogeneities (for example due to confinement, or background-charge densities, or designed impurity distributions) and memory effects, for example due to coherent feedback processes, phonon dephasing, or time-dependent current distributions. These systems will also display quantization effects.
TDDFT approaches based on local description and 3D reference systems have been shown to have problems with describing the 3D-2D crossover relevant for the quasi-2D quantum well-based systems \cite{Dagosta2007}, as we will discuss in Section \ref{sec:3.2.1}. Likewise, recent experiments \cite{Baboux2012} have pointed out
limitations of the SCD formalism to describe dissipation in these systems if a local approximation based on 3D reference systems is used (see Section \ref{sec:3.2.2}).

 \begin{figure*}
\includegraphics[width=\linewidth]{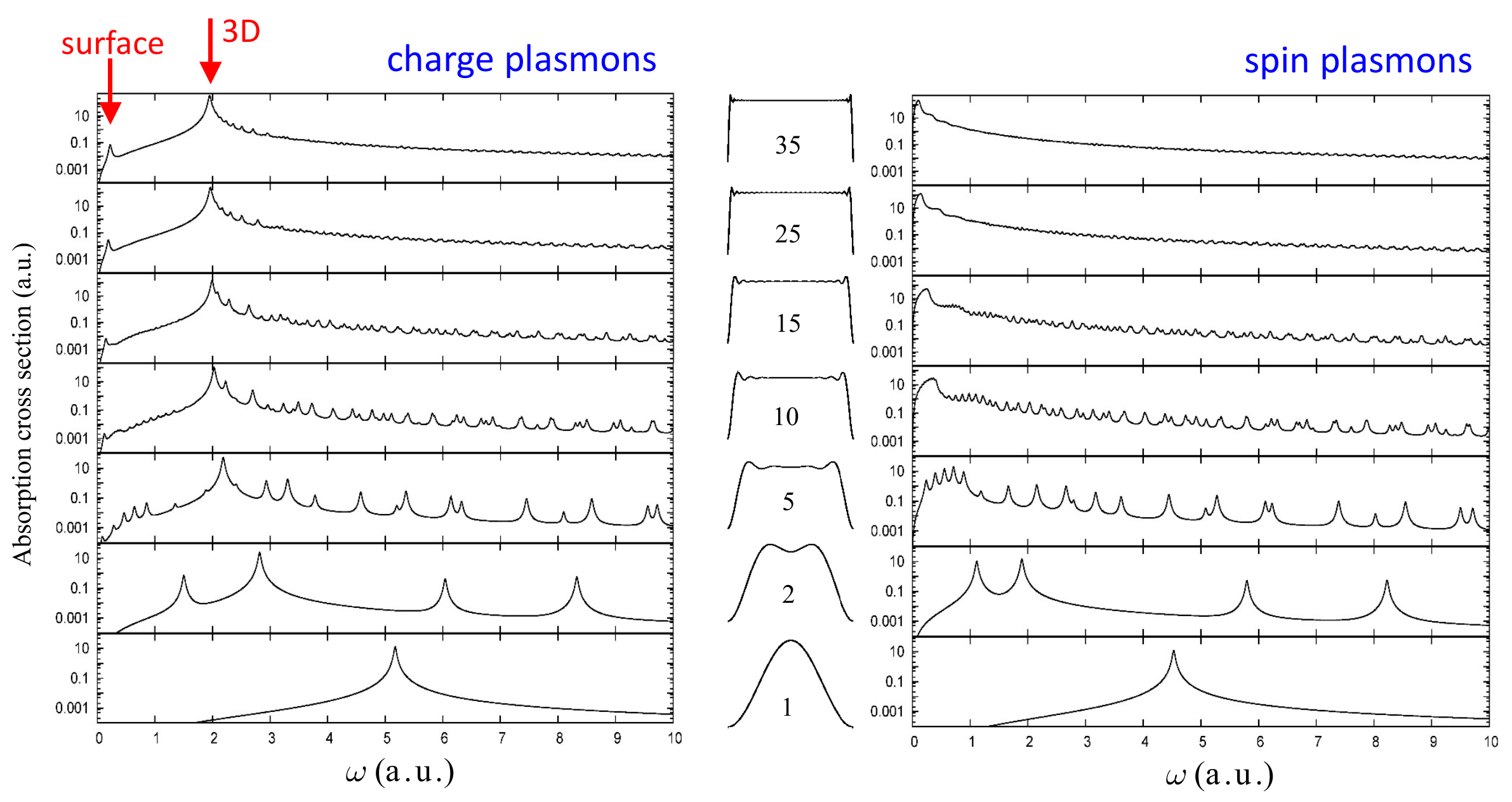}
\caption{Calculated absorption cross sections for $q=0$ intersubband charge and spin plasmon excitations in square quantum wells of increasing widths.
The insets show the quantum well density profiles at increasing numbers of occupied subbands (see text). The excitation spectra
evolve from the intersubband case, see middle panel of Fig. \ref{fig1}, to the 3D bulk case. The calculations were done with TDDFT using the 3D-ALDA,
not including SOC. Adapted from \cite{Karimi2014}.
} \label{fig8a}
\end{figure*}

\subsubsection{Excitation spectrum} \label{sec:3.2.1}
The many-body excitation spectrum of any system can, in principle, be exactly calculated using TDDFT within linear response \cite{Petersilka1995}. However, as the exact exchange-correlation kernel $f_{\rm xc}[n(\bfr,t)]$ defining the Kohn-Sham system is unknown, the accuracy of results will depend on the type of approximation used for this quantity \cite{Ullrich2012}. While $f_{\rm xc}$ is known to be a nonlocal functional of the electron density in both time and space, computationally not-too-demanding approximations usually assume locality or semi-locality. The simplest and most popular of these is the ALDA, which assumes locality in both time and space.

A relevant question is then if these (semi) local approximations are able to reproduce, at least qualitatively, the spectral features of 3D-2D crossover, in which nonlocality is assumed to play a strong role. A related question is up-to-which-point approximations based on the 3D electron liquid can be trusted in reproducing spectral features of quasi-2D systems, such as quantum wells.  These issues were systematically analysed in Ref. \cite{Karimi2014}.

The crossover from quasi-2D (i.e., quantum wells with a finite width) to 3D bulk-like is illustrated in Fig. \ref{fig8a}, which shows
ALDA intersubband excitation spectra at $q=0$, in the charge and spin channel, for quantum wells
with different subband occupation numbers. The well width and the sheet density $n_{\rm 2D}$
are chosen such that the average density remains constant at $\bar n = 0.30 \: {a_0^*}^{-3}$. However,
the density profile becomes more and more bulk like as the number of occupied subbands increases.

For a single occupied subband (the situation shown in Fig. \ref{fig1}, middle panels), the spectra show only a single peak, corresponding to the
intersubband charge and spin plasmons. As more subbands become occupied, more peaks show up, and eventually merge into very simple bulk
limits. The charge plasmon spectrum is then dominated by a single peak at
the bulk plasmon frequency $\omega_{\rm bulk}=\sqrt{4\pi \bar n}$, and there is also a smaller surface plasmon peak (red arrows in Fig. \ref{fig8a}).
On the other hand, the spin plasmon disappears in the 3D bulk limit, as expected.
Thus, the 3D ALDA correctly reproduces the physical features of the crossover from quasi-2D to 3D.

However, things are different in the opposite limit of increasingly narrow quantum wells, going from quasi-2D to strictly 2D.
In Ref. \cite{Karimi2014}, the performance of various xc kernels was compared for both inter- and intrasubband plasmon excitations,
and it was found that the 3D ALDA
breaks down and produces nonphysical results below a quantum well critical width $L_{\rm crit}^{\rm inter}\approx r_s$ for intersubband plasmons
and $L_{\rm crit}^{\rm intra}\approx 0.4r_s$ for intrasubband plasmons, where $r_s=1/\sqrt{\pi n_{\rm 2D}}$.
The relation $L_{\rm crit}^{\rm intra}<L_{\rm crit}^{\rm inter}$ implies that, in the limit of very narrow quantum wells, a 3D-based ALDA performs better for
describing in-plane than out-of-plane dynamics. Similar results were found for semilocal, gradient-corrected xc functionals.

It is worth noting that, for GaAs-based quantum wells and typical $n_{\rm 2D}$ values, the critical widths are relatively small, with e.g. $L_{\rm crit}^{\rm  inter}=17$ nm for $n_{\rm 2D}=10^{11}$ cm$^{-2}$. This is indeed good news given the popularity of ALDA. On the other hand, for systems in which these conditions are not met,
more sophisticated, nonlocal xc functionals (not based on the 3D electron liquid as reference system) should be used \cite{Karimi2014}.

\subsubsection{Intrinsic dissipation and linewidth of excitations}\label{sec:3.2.2}
For homogeneous systems the SCD can be phenomenologically introduced by writing the spin-drag friction force per unit volume exerted by the $\bsi$ spin population over the $\si$ spin population, moving with velocities $\bfv_\si$ and $\bfv_\bsi$, respectively \cite{Damico2000}:
\begin{equation} \label{Fho}
{\bf F}_{\si \bsi}^{\rm hom}(\omega) = e^2 n_\si n_\bsi  \Re \rho_{\si \bsi}^{\rm hom}(\omega,n_\si,n_\bsi) (\bfv_\si -\bfv_\bsi),
\end{equation}
where the spin-transresistivity $\rho_{\si \bsi}^{\rm hom}$ is a complex number, with its real part contributing to the drag coefficient \cite{Damico2000}.

For weakly inhomogeneous systems, one can consider the system as locally homogeneous and make a local approximation over the system volume $V$.
The power loss due to the SCD then becomes \cite{Damico2006}
\begin{eqnarray}
\lefteqn{{\cal P}_{\si V}(\omega) \approx  \int_V {\bf F}_{\si \bsi}^{\rm hom}(\omega; \bfr) \cdot \bfv_\si(\bfr)d\bfr }\nonumber \\
&=& e^2 \int_V  n_\si(\bfr) n_\bsi (\bfr) \Re \rho_{\si \bsi}^{\rm hom}(\omega,n_\si(\bfr),n_\bsi(\bfr)) \nonumber \\
& &{}\times [\bfv_\si(\bfr) -\bfv_\bsi(\bfr)]\cdot \bfv_\si(\bfr) d\bfr
.
\label{Pwinho}
\end{eqnarray}

Using a local-density approximation within linear-response TDDFT it is possible to derive, from first principles, an expression for the 2DEL intersubband spin-plasmon intrinsic linewidth \cite{Damico2000}. Very appealingly, this expression has a structure closely resembling that of the power loss, Eq.~(\ref{Pwinho}). However, when considering very narrow quantum wells (GaAs-based quantum wells, 20-25 nm wide), this approximation gives a linewidth of about 0.4 meV: comparison with experiments \cite{Pinczuk1989,Baboux2012} shows this to be an overestimate of the actual linewidth by about a factor 3.

There are two main issues with an approximation of the form of Eq.~(\ref{Pwinho}) when considering a narrow quantum well and a plasmon whose associated spin-current is in the growth direction. First, is the approximation good enough to account for the quantum well's strong inhomogeneity in the growth direction? Second, which are the consequences of using a 3D reference system for constructing the local approximation?

To answer these questions, a fully inhomogeneous theory of SCD was developed in Ref.~\cite{Damico2013} and then applied to the case of the intersubband spin plasmon of a quantum well. Eq.~(\ref{Fho}) is generalized to the fully inhomogeneous microscopic expression
\begin{eqnarray} \label{F2}
{\bf F}_{\si \bsi}(\bfr,\bfr',\omega) &=& e^2 n_\si(\bfr) n_\bsi(\bfr')  \Re \stackrel{\leftrightarrow}{\rho}_{\si \bsi}(\bfr,\bfr',\omega)
\nonumber\\
&&\times \left[\bfv_\si(\bfr) -\bfv_\bsi(\bfr') \right],
\end{eqnarray}
where $\stackrel{\leftrightarrow}{\rho}_{\si \bsi}(\bfr,\bfr',\omega)$ is now a non-local tensor.
Taking into account the homogeneity in the in-plane directions and the strong inhomogeneity in the growth ($z$) direction in a quantum well, the corresponding intersubband spin plasmon linewidth becomes
\begin{eqnarray} \label{Gnonloc}
\Gamma_{\rm SCD}^{\rm nonloc} &=& \frac{e^2 n_{\rm 2D}}{2m^* \Omega_s}\int dz\int dz' \:
n_\si(z)n_\bsi(z')\nonumber\\
&&\times \Re \rho_{\ua\da}^{zz}(q=0,z,z',\Omega_s)
\nonumber\\
&& \times  [v_{12}^2(z) + v_{12}(z)v_{12}(z') ],
\end{eqnarray}
where $\Omega_s$ is the spin-plasmon frequency at $q=0$, and $v_{12}(z)$ is the velocity profile of the spin-plasmon mode.
The quasi-2D dimensionality of a narrow quantum well is accounted for by constructing $\Re \rho_{\ua\da}^{zz}$ in a mixed $(\bfq,z,z')$ representation  \cite{Damico2013}.

In contrast to the local approximation (\ref{Pwinho}), $\Gamma_{\rm SCD}^{\rm nonloc}$ strongly depends on the quantum well features, and accounts for the strong inhomogeneity and the quantization in the growth direction. All this has implications for the allowed processes for the decay of the spin plasmon: a large momentum parameter-space region is now forbidden, and wide regions corresponding to strong Coulomb interaction cannot contribute. As a result the SCD becomes much less effective and the plasmon linewidth is drastically reduced; the estimate for the intrinsic linewidth for narrow GaAs-based quantum wells is now of the order of 0.02-0.01 meV, about 15-20\% of the experimental results \cite{Pinczuk1989,Baboux2012}.

Additional contributions to the intersubband spin-plasmon linewidth will come from extrinsic (e.g. impurities and surface roughness) and {\em mixed} intrinsic-extrinsic contributions to the spin-transresistivity. In fact it can be shown \cite{Damico2013} that the relevant spin-transresistivity tensor component, when derived from the generalized Kubo formula, is given by
\begin{equation} \label{12}
\Re \rho^{zz}_{\si \bsi}(\bfr,\bfr',\omega) =
\frac{m^2}{\omega e^2} \frac{
\Im \langle \langle \dot J^z_{\si}(\bfr) ; \dot J^z_{\bsi}(\bfr')\rangle \rangle_\omega}
{n_\si(\bfr) n_{\bsi}(\bfr')} \:,
\end{equation}
where
$\dot J^z_{\si}(\bfr) = -\frac{i}{\hbar}[J^z_\sigma, H]$,
and $H$ is the many-body Hamiltonian of the system containing kinetic, Coulomb ($W_C$), and external potential ($V_{ext}$) terms. Contributions from the mixed terms combining $[J^\alpha_\sigma, W_C]$ and $[J^\alpha_\sigma,V_{ext}]$ commutators can be estimated for scattering from remote impurities ($\delta$-layer doping) \cite{Da-Ul-unp}. Results show that these terms  contribute about 0.005 meV to the linewidth, suggesting that the dominant contribution to the measured linewidth comes from surface roughness and/or inhomogeneous broadening.

\subsection{Chiral spin waves}\label{sec:3.3}
The concept of Chiral Spin Resonance (CSR) was introduced in \cite{Shekhter2005} to indicate the $q=0$ resonant transition between electron states split by SOC in a 2DEL with no applied static magnetic field and driven by a high-frequency electromagnetic field. SOC couples the directions of electronic spin and momentum, so that this resonance connects  states with opposite chirality. Because under SOC the electron spin is not conserved, the width and frequency of the CSR is renormalized by electron-electron  interactions \cite{Shekhter2005}.

The corresponding long-wavelength regime was analyzed in  \cite{Ashrafi2012}, with the prediction of collective modes termed  `chiral spin waves', corresponding to in and out-of-plane modulations of the magnetization.  In the ballistic limit,  the stiffness of their dispersion relation is affected by the strength of the electron-electron  interaction; the lower energy mode is the out-of-plane (longitudinal) spin-chiral wave, which merges with the particle-hole continuum at $q v_{\rm F} = 2 |\alpha| k_{\rm F}$.  For small $q$ values and  strong enough electron-electron interaction, the in- and out-of-plane modes could have opposite curvature. By lateral modulation of the SOC in the 2DEL plane, standing chiral spin waves could be generated and experimentally observed.

In a related study \cite{Maiti2015a}, the intrinsic linewidth of the chiral spin waves due to the interplay of SOC and momentum exchange between `$+$' and `$-$' spin populations (transversal SCD) was analyzed using diagrammatic expansion techniques. It was shown that, because of SOC, dissipation due to transverse SCD -- usually vanishing as $q\to0$ -- remains non-zero even at $q=0$. The damping rate is proportional to the square of the SO splitting renormalized by the Fermi energy.

The observation of chiral spin waves may have been realized in 2DELs confined in GaAs quantum wells. Indeed in Ref.~\cite{Jusserand1995}, the SO-split particle-hole continuum is clearly observed with ERRS (see Section \ref{sec:2.7}), together with an additional peak, which at the time of the publication of Ref. \cite{Jusserand1995} was not understood. The SOC strength in GaAs is weak and renders the confirmation of chiral spin waves very difficult as their energy is very close to the energy cut-off of the Raman technique.

\begin{figure*}
\includegraphics[width=\linewidth]{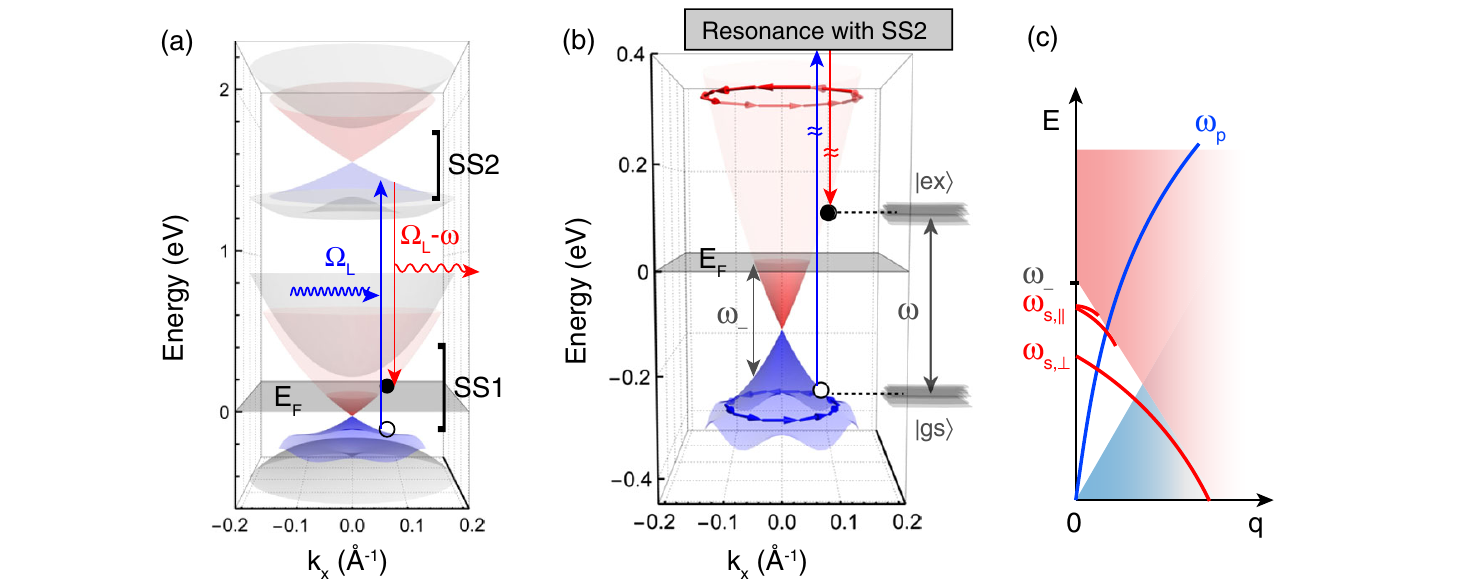}
\caption{(a) Band structure of the surface states of Bi$_2$Se$_3$, showing Dirac cones of opposite chirality in blue and red.
The arrows illustrate resonant Raman process (via the unoccupied surface states SS2) that were used in the experiment.
(b) Close-up view of the region around the Fermi level, where $\omega_-$ is the threshold energy for transitions between
the two Dirac cones. (c) Schematic illustration of the particle-hole continua in the charge (blue) and spin (red) channels.
The charge plasmon dispersion is in blue, and the chiral spin modes are in red.
\copyright 2017 American Physical Society. Reprinted, with permission, from \cite{Kung2017}.
}
\label{fig9}
\end{figure*}

Very recently, the zone center chiral spin wave was observed by ERRS in the helical 2DEL which forms in the topological bands of Bi$_2$Se$_3$ \cite{Raghu2010,Kung2017}. In this material, the topological surface states lying in the bulk gap form helical Dirac cones (see Fig. \ref{fig9}). When the doping level is finely adjusted in this gap, the situation becomes close to a 2DEL with a giant SOC, except that the kinetic energy is linear in momentum. For a full demonstration of the existence of chiral spin waves in this system, the mode dispersion at finite wavevectors remains to be observed.

\section{Spin-polarized 2DEL}\label{sec:4}

The spin-polarized 2DEL (SP2DEL) originally attracted a large number of experimental and theoretical investigations of its ground state properties. In fact, a prediction \cite{Ando1982,Attaccalite2002,Varsano2001} that a spontaneous spin polarization should occur at low density ($r_s\simeq2.3$) due to Coulomb-exchange (see Section \ref{sec:2.1}), seemed to be in conflict with the Mermin-Wagner theorem (ferromagnetism is forbidden in 2D). This was also connected with the mystery of the metal-insulator transition discovered at $r_s=8$ in Si inversion layers \cite{Pudalov02,Shashkin06} and more generally stimulated developments of spin-resolved formalisms for Coulomb many-body phenomena \cite{Ryan1991a,Polini01,Marinescu97,Yi96,Rajagopal78}.

The SP2DEL is a somewhat idealized system where the spin-polarization  degree $\zeta$ is finite (see Section \ref{sec:2.1}),  caused by an external magnetic field, but it must be without Landau quantization. In contrast to earlier studies of spins in GaAs/GaAlAs systems, where Landau orbital quantization dominated over spin quantization \cite{Pinczuk1992,Davies1997}, such an ideal object is not easy to obtain: the external magnetic field must be applied in the plane of the 2DEL and its magnetic length $l_m$ (see Section \ref{sec:2.6}) must be lower than the characteristic 2DEL thickness. Nevertheless, to achieve high $\zeta$, one necessarily breaks the latter condition at some magnetic field, leading to the formation of magneto-hybrid subbands in the 2DEL. The doped diluted magnetic quantum well detailed below allowed a new approach to this problem.

\subsection{Experimental model system}\label{sec:4.1}
\begin{figure}
\includegraphics[width=\linewidth]{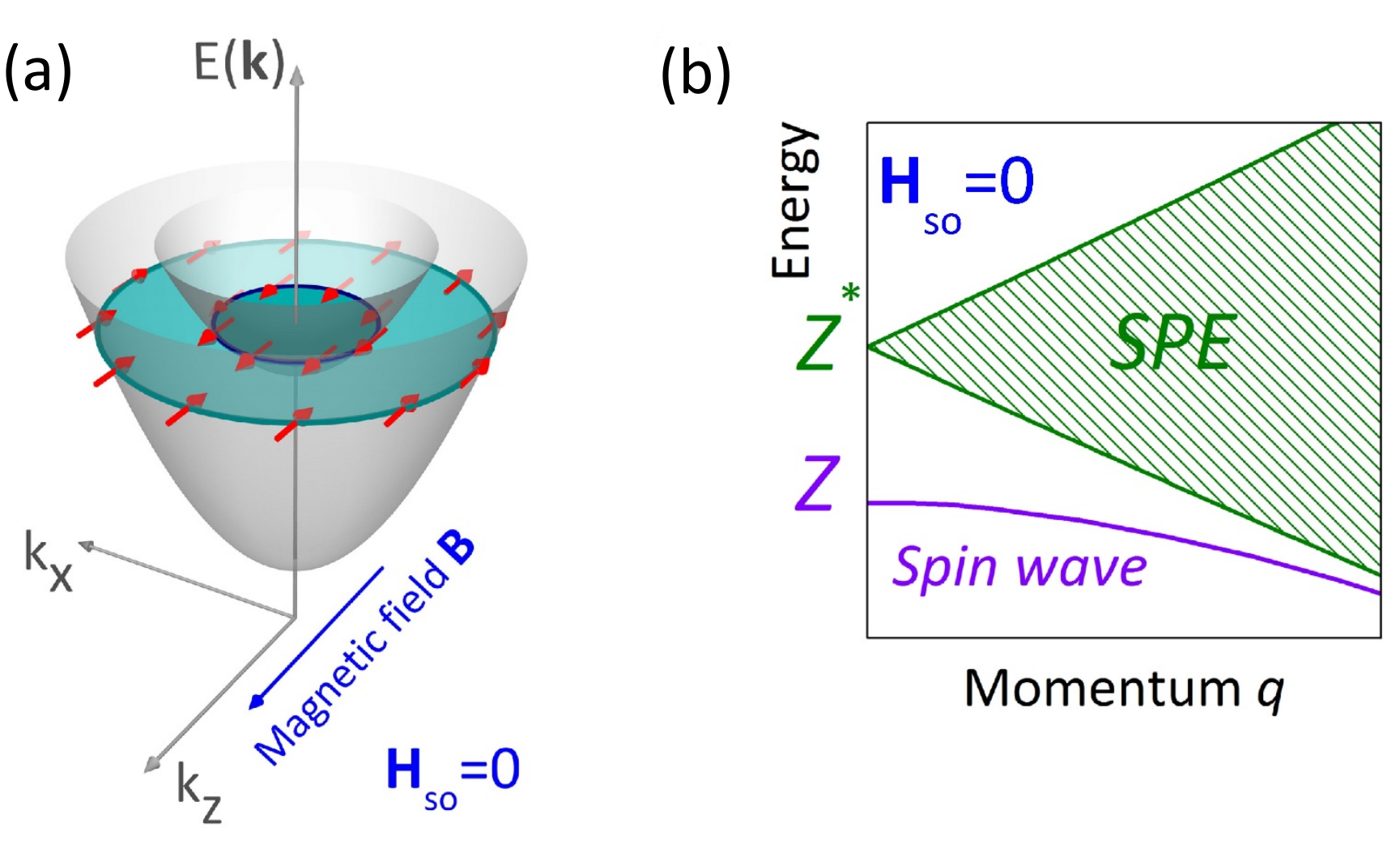}
\caption{(a) Spin-split parabolic subbands of the 2DEL in the absence of SOC. The external magnetic field $\bfB=B_{\rm ext}\bfz$ is applied along the $z$-axis lying in the plane of the quantum well. Conduction states are filled up to the Fermi energy. The spin `up' (minority) and spin `down' (majority) Fermi disks are highlighted. (b) Spin-excitation spectrum without SOC: spin waves propagate in the energy gap below the single-particle spin-flip excitation (SPE) continuum. $Z^\ast$ is the zone center SPE energy \cite{Perez2007}. $Z$ is the homogenous spin wave energy.}
\label{fig10}
\end{figure}

A 2DEL embedded in a high mobility Cd$_{1-x}$Mn$_x$Te quantum well \cite{Karczewski1998} was successfully introduced in 2003 as a test bed for the SP2DEL \cite{Perez2009,Perez2007,Jusserand2003}. In Cd$_{1-x}$Mn$_x$Te quantum wells, the exchange coupling between the 2DEL ($s$-electrons) and $d$-electrons of Mn impurities introduces a Zeeman energy $Z$ which is controlled by $x$. The form of $Z$ is given in Eq. (\ref{ZB0}) and can be recast in $\hat H_{\rm m}$ of Eq. (\ref{H}) as an external magnetic field of the form $\bfB_{\rm sd}=(Z/g^*\mu_B)\hat e_{\bfB_{\rm ext}}$. The key value is the energy $J_{sd}N_{\rm Mn}=x_{\rm eff}\times 220 \,{\rm meV}$, see Eq.~(\ref{Delta}), where $x_{\rm eff}\leq x$ is the unpaired spin number per unit cell \cite{Gaj1979}, and $J_{sd}\simeq 14.96 \,{\rm meV \, nm}^3$ and $N_{\rm Mn}\simeq x_{\rm eff}\times 14.70{\rm nm}^3$ are band structure dependent quantities. Hence, for $x_{\rm eff}=1\%$, one finds
$Z=2.2 \,{\rm meV}$, which is already equivalent to the substantial magnetic field strength of $|\bfB_{\rm sd}|\simeq 27{\rm T}$. Thus, as depicted in Fig. \ref{fig10}, the individual states of the SP2DEL are divided into two spin-split subbands occupied up to the Fermi energy. In the absence of $\hat H_{\rm SO}$, the equilibrium spins are antiparallel in the 2DEL plane.

\subsection{Excitations of the SP2DEL} \label{sec:4.2}

Similar to the unpolarized case, the excitation spectrum of the SP2DEL is divided into single-particle and collective excitations, see Fig. \ref{fig10}. Both are subdivided into longitudinal and transverse type, depending on whether they involve spin-conserving or spin-flip processes.

The main feature of the SP2DEL is the opening of a gap in the energy-momentum excitation spectrum, which allows a collective spin wave to propagate separately from the single-particle excitations. As introduced in Section \ref{sec:2.6.2}, we recall the spin-wave operator $\hat{S}_{\mathbf{q}}^+=\hat{S}_{x,\mathbf{q}}+i\hat{S}_{y,\mathbf{q}}=\sum_{\mathbf{k}}c_{\mathbf{k-q},\uparrow}^{+}c_{\mathbf{k},\downarrow}^{\vphantom{+}}$, where $c_{\mathbf{k},\uparrow}^{+}$ and $c_{\mathbf{k},\downarrow}^{\vphantom{+}}$ are electron creation-anihilation operators. A spin-flip (transverse)
individual excitation (SF-SPE) of the spin-polarized ground state $|0\rangle$ is simply $c_{\mathbf{k-q},\uparrow}^{+}c_{\mathbf{k},\downarrow}^{\vphantom{+}}|0\rangle$, where an electron of momentum $\mathbf{k}$ and spin $\downarrow$ is promoted to the empty state $\mathbf{k-q}$,$\uparrow$.

\subsubsection{Dynamics of single-particle excitations} \label{sec:4.2.1}

As stated, the SP2DEL Hamiltonian is the one of Eq. (\ref{H}) without $\hat H_{\rm SO}$. The kinetic part is $\hat{H}_{\mathrm{K}}=\sum_{\mathbf{k,\sigma}}E_{\mathbf{k}%
}c_{\mathbf{k},\sigma}^{+}c_{\mathbf{k},\sigma}^{\vphantom{+}}$, $\hat{H}_{\mathrm{C}}$ is the Coulomb part and $\hat H_{\rm m}=Z\hat{S}_{z,\mathbf{q}=0}$ is the Zeeman part.
The individual modes are conserved by $\hat{H}_{\mathrm{K}}$ and $\hat H_{\rm m}$. $\hat{H}_{\mathrm{C}}$ has a part that directly acts on the particle-hole pairs, and a remaining part which couples them to multi-electron-hole pairs. The former renormalizes $Z$ into $Z^*$ (see Section \ref{sec:2.6.1}). The latter has several consequences:
it is the origin of the transverse SCD (see Section \ref{sec:2.3})  and can be described by an
electron-electron scattering time  $\tau_{e-e}$ \cite{Hankiewicz2008};
$\hat{H}_{\mathrm{C}}$ conserves the global spin, which means that multi-electron-hole pairs are the product of spin one and spin zero pairs, where the latter are the elemental components of longitudinal collective modes \cite{Gomez2010}.

The equation of motion for SF-SPE reads
\begin{eqnarray}\label{SFSPE}
\big[&&c_{\mathbf{k-q},\uparrow}^{+}c_{\mathbf{k},\downarrow}^{\vphantom{+}},\hat{H}_{\mathrm{K}}+\hat{H}_{\mathrm{C}}+\hat{H}_{\mathrm{m}}\big]  = \nonumber\\%
&&\left(  E_{\mathbf{k}}-E_{\mathbf{k-q}}-Z^*+i\frac{\hbar}{\tau_{e-e}}\right)c_{\mathbf{k}%
-\mathbf{q},\uparrow}^{+}c_{\mathbf{k},\downarrow}^{\vphantom{+}} \nonumber\\
&&+\sum_{\rm Spin 1}\text{Multi pairs} .
\end{eqnarray}
Compared to the bare Zeeman energy $Z$, $Z^*$ is enhanced by Coulomb-exchange between spin-polarized electrons, a phenomenon linked to the
spin-susceptibility enhancement \cite{Perez2007}. SF-SPEs are characterized
by two wavevectors, $\mathbf{k}$ and $\mathbf{q}$, they are degenerate to $Z^*$ at $\mathbf{q}=0$ and form a continuum when $\mathbf{q}\neq0$ (see Fig. \ref{fig10}).

\subsubsection{Spin waves} \label{sec:4.2.2}

\begin{figure}
\includegraphics[width=\linewidth]{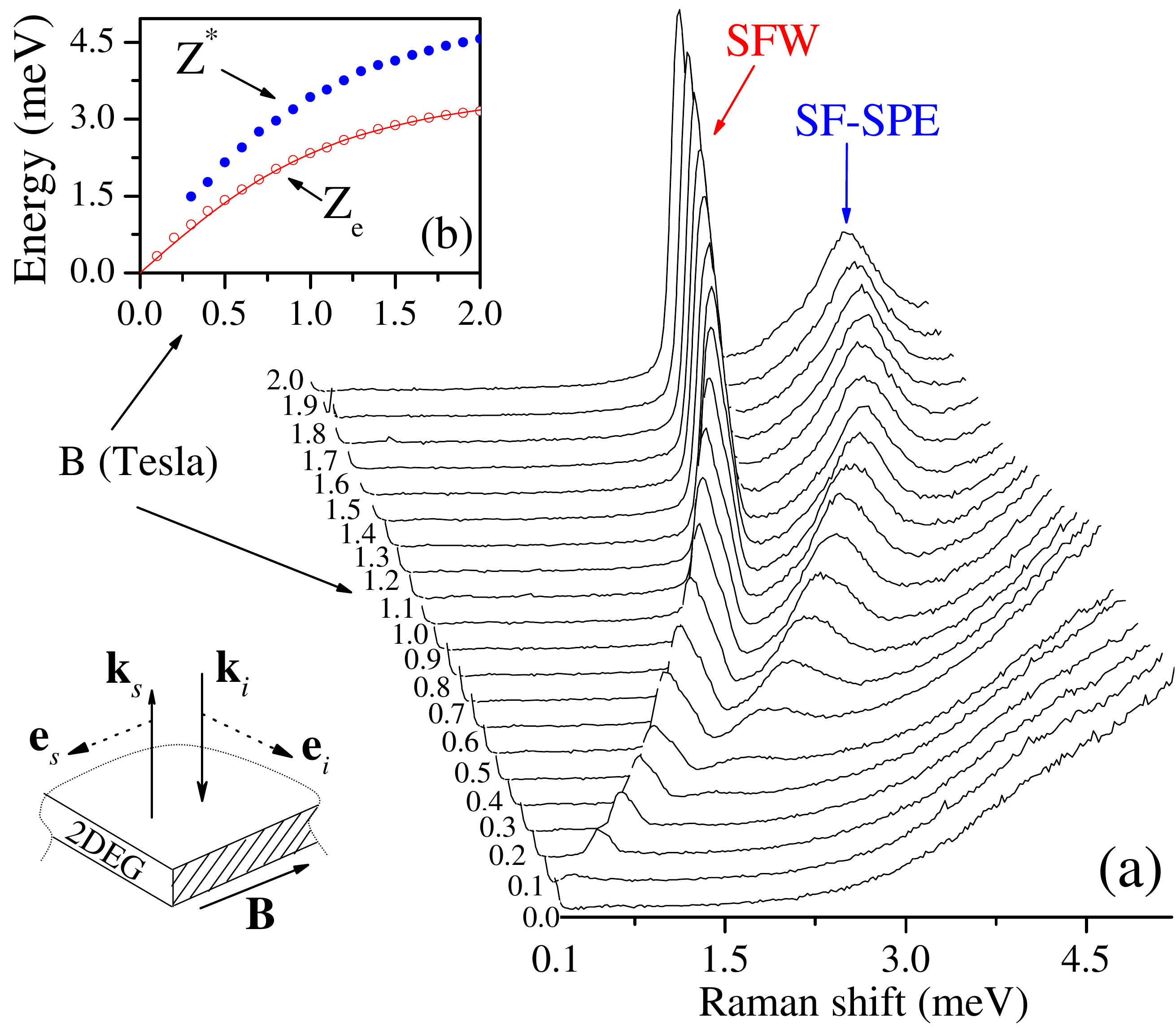}
\caption{\textbf{(a)} Cross-polarized Raman spectra with incident and scattered beam along the growth axis of the quantum well (see inset), taken for various values of in-plane magnetic field $B_{\rm ext}$.  Spin-flip excitations are probed at $q=0$. The low energy line is the spin-flip wave (SFW), the other signal represents the spin-flip single-particle excitations (SF-SPE). \textbf{(b) }Peak positions of the SFW lines (open circles) and SF-SPE (full circles) as a function of the magnetic field. The SFW peaks are fitted with Eq. (\ref{ZB0}) to obtain the Mn concentration $x=0.75$\% and the electronic temperature $T=1.5{\rm K}$.
\copyright 2007 American Physical Society. Reprinted, with permission, from \cite{Perez2007}.}
\label{fig11}
\end{figure}

Since $\hat{H}_{\mathrm{C}}$ conserves the macroscopic spin, the Coulomb interaction makes no contribution
in the equation of motion of the spin-wave operator $\hat{S}_{\mathbf{q}}^+$:
\begin{equation}\label{SW}
\big[  \hat{S}^+_{\mathbf{q}},\hat{H}_{\mathrm{K}}+\hat{H}_{\mathrm{C}}+\hat{H}_{\mathrm{m}}\big]  =-Z\hat{S}^+_{\mathbf{q}}+\hbar\mathbf{q\cdot
\hat{J}}^+_{\mathbf{q}} .
\end{equation}
The second term on the right-hand side is the transverse spin-current operator, $\mathbf{\hat{J}}^+_{\mathbf{q}}=\frac{\hbar}{m^{\ast}}\sum_{\mathbf{k}%
}\left(  \mathbf{k-}\frac{\mathbf{q}}{2}\right)c_{\mathbf{k-q},\uparrow}^{+}c_{\mathbf{k},\downarrow}^{\vphantom{+}}$. Equation (\ref{SW}) has several consequences. On one hand, for $\bfq=0$, the state $\hat{S}^+_{\mathbf{q}=0}|0\rangle$ is an exact eigenstate of the SP2DEL whose excitation energy is exactly $Z$. This means that, despite the
fact that this state describes a collective motion where spins precess in phase, its precession frequency $Z/\hbar$ has no contribution from the Coulomb interaction. This exact result is called the \emph{Larmor Theorem}:

\begin{equation}\label{Larmor}
    \frac{d}{dt}\hat{S}^+_{\mathbf{q}=0}=i\frac{Z}{\hbar}\hat{S}^+_{\mathbf{q}=0}.
\end{equation}
The Coulomb interaction affects instead, in the $\bfq=0$ limit, the SF-SPE precession frequency $Z^*/\hbar$. ERRS spectra nicely evidence these two excitations, as can be seen in Fig.~\ref{fig11}.

On the other hand, the spin-wave dispersion results from the spin current. It is an interplay between the motion in a parabolic band and Coulomb interaction. However, the spin current (which is a superposition of SF-SPE with different velocities) has an equation of motion similar to Eq. (\ref{SFSPE}). Thus it will also introduce a coupling with longitudinal modes and an intrinsic damping due to SCD, as we discussed earlier in Section \ref{sec:3.2.2}.
The dispersion and damping of the spin wave can be found with linear response theory \cite{Perez2009,Perez2011,Gomez2010}; in summary, one obtains
\begin{equation}
\frac{d}{dt}\hat{S}^+_{\mathbf{q}}=i\left(\omega_{q}+i\eta_{q}\right)\hat
{S}^+_{\mathbf{q}},
\end{equation}
\begin{equation}
\omega_{q}=Z/\hbar-\sws\frac{\hbar}{2m^{\ast}}q^{2}\label{ReOmegaq},
\end{equation}
\begin{eqnarray}
\eta_q &=& q^{2}\frac{\hbar}{2m^{\ast}\left\vert
\zeta\right\vert }\frac{3 Z^{\ast}\eta_\mathrm{sp}}{\left(  Z^{\ast}\right)
^{2}+\eta_\mathrm{sp}^{2}} \nonumber\\
&&\times \left[  \frac{Z^{\ast}}{Z}-\frac{\left(  Z^{\ast
}\right)  ^{2}+\frac{1}{3}\eta_\mathrm{sp}^{2}}{\left(  Z^{\ast}\right)  ^{2}+\eta_\mathrm{sp}^{2}%
}\right], \label{ImOmegaq}
\end{eqnarray}
where $\sws=\frac{1}{\left\vert
\zeta\right\vert }\frac{Z}{Z^{\ast}-Z}$ is the spin-wave stiffness and $\eta_\mathrm{sp}$ is the SF-SPE scattering rate.
We note that both disorder and transverse SCD would contribute with the same $q^2$ dependence to   $\eta_q$ \cite{Hankiewicz2008}.

\begin{figure}
\includegraphics[width=\linewidth]{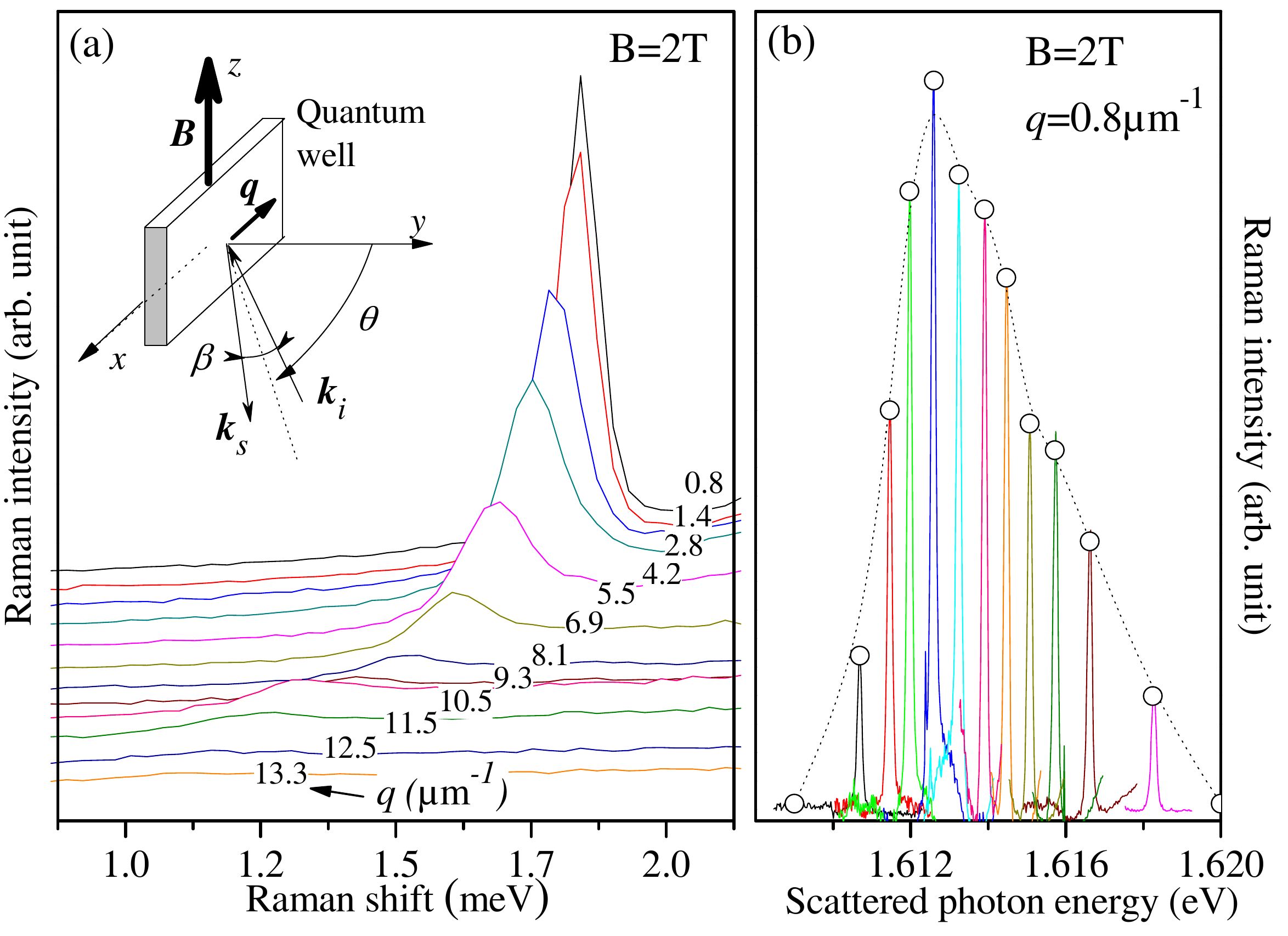}
\caption{\textbf{(a)} Typical cross-polarized Raman spectra obtained
at $B_{\rm ext}=2{\rm T}$ and for different values of $q$. The single Raman line
is the SFW. Inset:  scattering geometry showing the definition of the angles.
Incoming photon is polarized parallel to $\bfB_{\rm ext}$ ($\pi$), while the
scattered one is polarized perpendicular to $\bfB_{\rm ext}$ ($\sigma$). \textbf{(b)} Spectra
obtained by shifting the laser wavelength. Amplitude variations of the Raman
line reveal the optical resonance width.
\copyright 2010 American Physical Society. Reprinted, with permission, from \cite{Gomez2010}.}%
\label{fig12}
\end{figure}

Evidence of the universal $q^{2}$-laws presented in Eqs.~(\ref{ReOmegaq}) and (\ref{ImOmegaq}) has been provided in high
mobility SP2DELs, as described in this section. Since the well-defined spin-wave modes have
been successfully observed in these quantum wells \cite{Perez2007,Jusserand2003}, this
material is a perfect candidate to investigate these laws. The sketch in
Fig. \ref{fig12} depicts the experimental geometry: the external
magnetic field, $\bfB_{\rm ext}$, is applied in the $z$ direction parallel to the
quantum well plane and the average angle $\theta$ of the incoming and
back-scattered light wavevectors with respect to the normal direction can be
tuned to make the in-plane Raman transferred wavevector $q=\frac{4\pi}%
{\lambda}\cos\frac{\beta}{2}\sin\theta$ vary in the range $0<q<16 \mu{\rm m}^{-1}$,
$\beta\simeq5^{\circ}$, and $\lambda$ is the incoming light wavelength.

In Fig. \ref{fig12}a, cross-polarized Raman
spectra are plotted. They are obtained for increasing $q$ and fixed external magnetic field at
superfluid He bath temperature ($\mathrm{T}\sim 2.0{\rm K}$). These spectra present
a clear dispersive Raman line associated to the spin wave. The resonant behavior of the Raman peak is shown in Fig. \ref{fig12}b. Tuning the laser wavelength
across the optical resonance evidences a resonance width which is 20 times
larger than the SFW Raman line. Hence, we can consider that
Raman spectra give access to $\Im\chi_{\da\ua,\da\ua}(\bfq,\omega)$ and extract from these data both the \SWd energy ($\hbar\omega_{\mathrm{sw}}$)
and the $q$-dependence of the linewidth $\eta_{\mathrm{sw}}$.

\begin{figure}
\includegraphics[width=\linewidth]{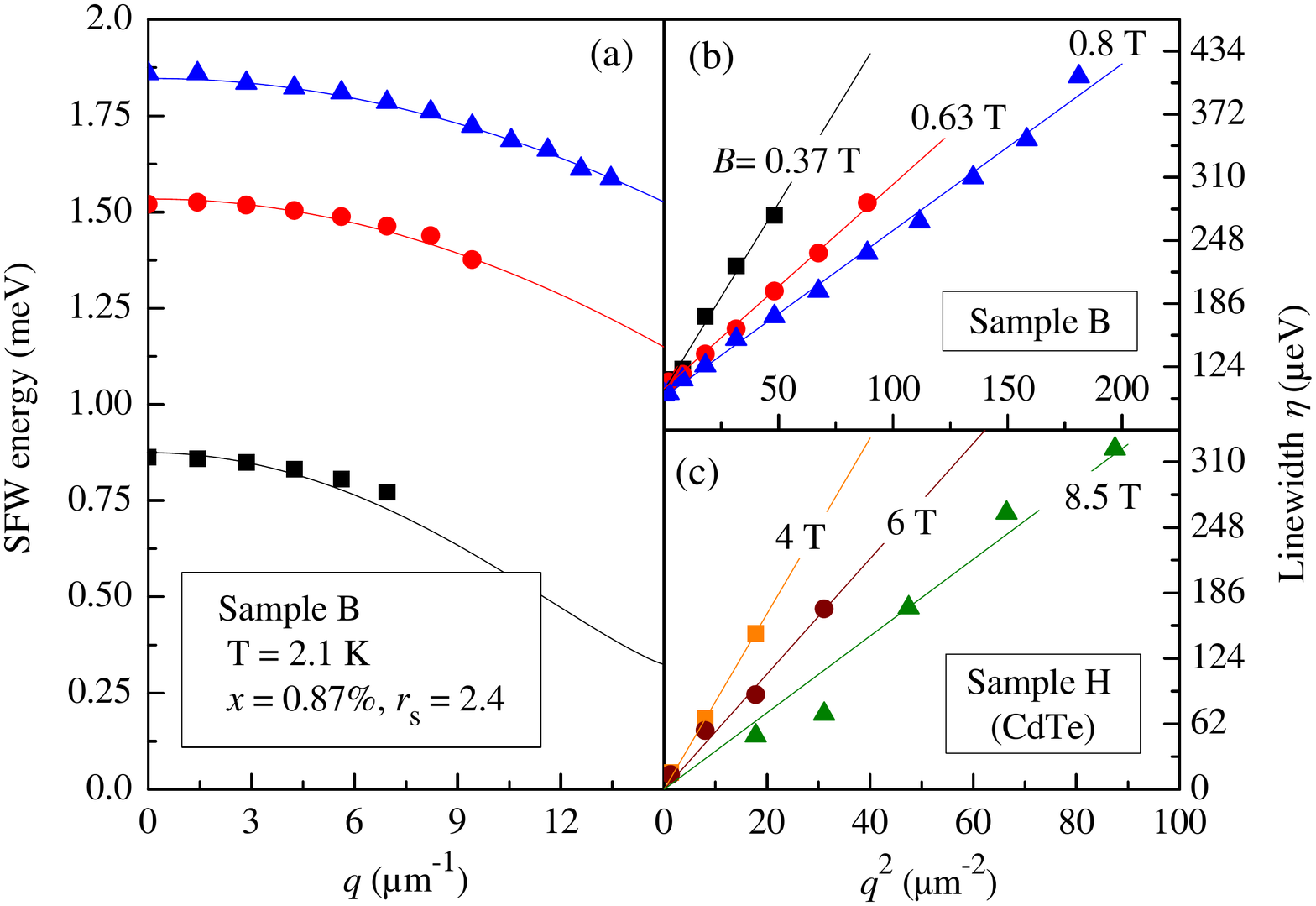}
\caption{ Typical \SWd energy \textbf{(a)} and linewidth \textbf{(b)}
$q$-dependence obtained on a sample with parameters $x=0.87$ and $r_{s}=2.4$, for $B_{\rm ext}$=0.37, 0.63
and 0.8 T. In agreement with Eqs. (\ref{ReOmegaq}) and (\ref{LineWidthq2}) the
data follow a parabolic behavior. \textbf{(c)} Linewidth $q$-dependence obtained on a CdTe
sample (without Mn). \copyright 2010 American Physical Society. Reprinted, with permission, from \cite{Gomez2010}.}%
\label{fig13}
\end{figure}

As shown in Fig.~\ref{fig13}a, $\hbar\omega
_{\mathrm{sw}}$ is well reproduced by the
formula of Eq. (\ref{ReOmegaq}). Extraction of the widths of the Raman lines needs an accurate deconvolution process with the spectrometer response \cite{Gomez2010};
the results are plotted in Fig. \ref{fig13}b as a function of $q^{2}$ for the
same conditions as the dispersions plotted in Fig. \ref{fig13}a.
As disorder effects dominate over the transverse SCD, here the latter is neglected and the scattering time is assumed to be due to disorder only \cite{Gomez2010}.
It is found that, in the explored range of wavevectors $\left(q\ll k_{\rm F}\right)
$, the linewidth and magnetic field $q$  dependencies are very well reproduced by the parabolic form
\begin{equation}
\eta_{\mathrm{sw}}=\eta_{0}+\eta_q=\eta_0+\eta_{2}q^{2},
\label{LineWidthq2}
\end{equation}
where $\eta_{0}$ is necessary to account for the homogeneous mode ($q=0$)
damping caused by any source that breaks the \emph{Larmor Theorem}: here, Mn spin fluctuations. Indeed, these are known to introduce a strong
damping in the homogeneous mode \cite{Crooker1996}. In
the CdMnTe quantum wells, the typical Mn average distance $\bar{d}\sim0.4$ nm is far smaller
than the minimum magnetization wavelength probed in the Raman experiment
($q\bar{d}\ll1).$ Hence, Mn damping is expected to be constant in the explored
range of $q$ and contributes to $\eta_0$ but not to $\eta_{2}$.  Fig. \ref{fig13}c confirms the presence of the $q^2$ law
with the same order of magnitude in a CdTe quantum well (without Mn).

\begin{figure}[]
\centering
\includegraphics[width=\columnwidth]{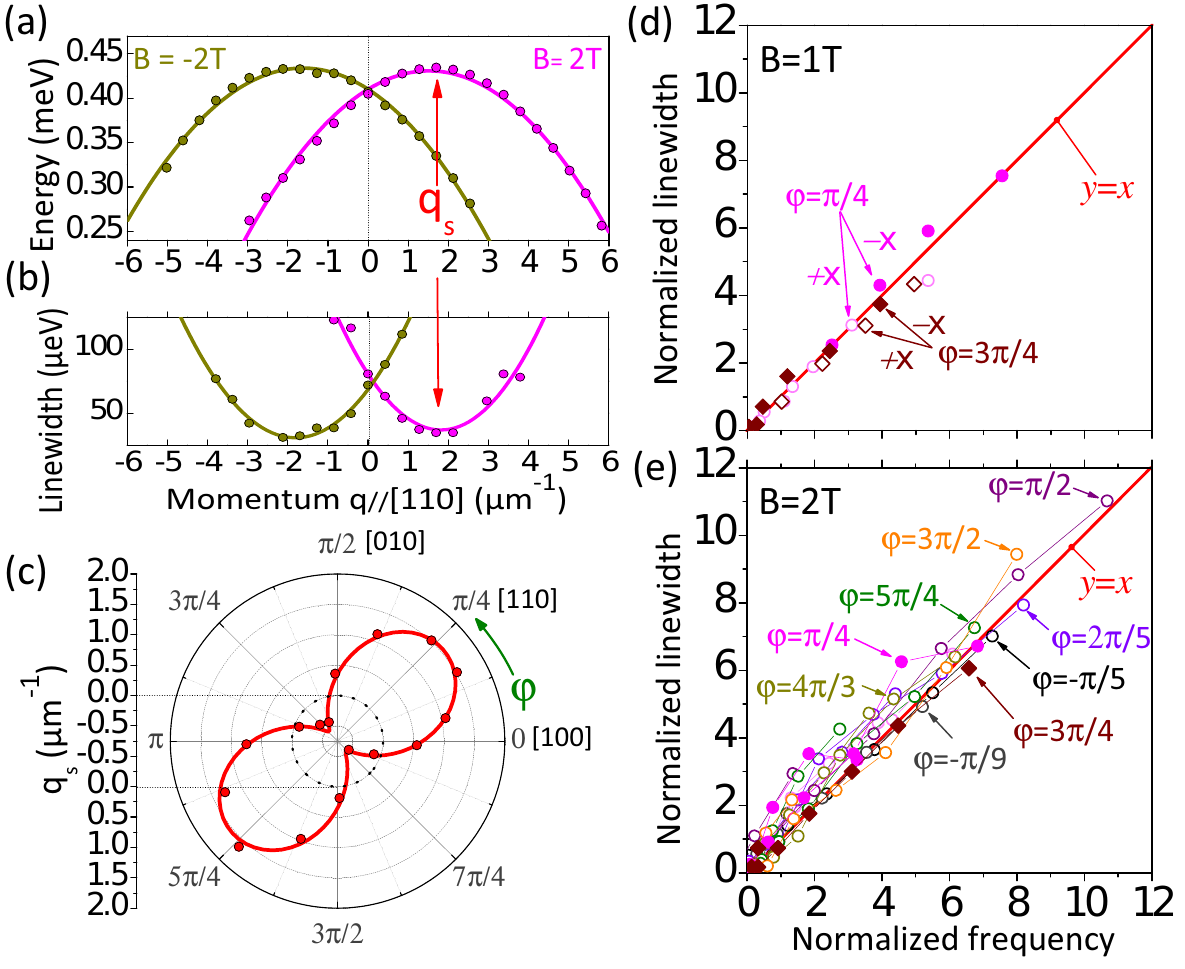}
\caption{(a\&b) Momentum dispersion of energy (a) and linewidth (b) of the \SW for the in-plane direction $\varphi= \pi /4$ and $B_{\rm ext}=\pm 2{\rm T}$. Dispersions are shifted by $q_s$ from $q=0$ with a mirror symmetry when inverting the magnetic field, see Eq. (\ref{ChirOmega}).  (c) ({\Red \textbullet}) represents the $q_{s}$ dependence with $\varphi$, extracted from the measured dispersions. The red curve is a fit with the theoretical value of $q_s(\varphi)$ (see Ref. \cite{Perez2016}). (d\&e) Universal linear relation between the linewidth and the energy of the spin wave: $(\eta-\eta_0)/\eta_2$ is plotted as a function of $\frac{2m^\ast}{\hbar^2}(\hbar\omega-Z)/\sws$, symbols of the same color are for a given in-plane angle $\varphi$, but for various values of $q$. (d) $B_{\rm ext}=+1{\rm T}$, open (solid) symbols correspond to spin waves with wavevector $\bfq=q\bfx$ directed towards $-\bfx$ ($+\bfx$). (e) $B_{\rm ext}=+ 2{\rm T}$, solid symbols correspond to the two extremal angles $\varphi=\frac{\pi}{4},\frac{3\pi}{4}$, open symbols are for other angles.
\copyright 2016 American Physical Society. Reprinted, with permission, from \cite{Perez2016}.}
\label{fig14}
\end{figure}

\subsection{Chiral Spin Waves} \label{sec:4.3}

Similarly to the unpolarized case of Section \ref{sec:3.3}, chiral spin waves exist and have been successfully observed in the model system of section~\ref{sec:4.1}. SOC of the conduction band is at the origin of the chirality. In reality, SOC is always present in asymmetrically doped quantum wells, but what matters here is the relative importance of the Coulomb strength, Zeeman energy and SOC. To successfully evidence the chiral spin waves in the above SP2DEL, a required condition is $\alpha k_{\rm F}\sim Z\sim Z^*-Z$. The former is the typical strength of SOC as introduced in Eqs. (\ref{Rashba}) and (\ref{Dresselhaus}), the second is the Zeeman energy and the latter is the Coulomb-exchange-correlation strength. This condition was met in Ref. \cite{Perez2016} by adjustment of the quantum well width, 2DEL density $n_{\rm 2D}$ and Mn concentration $x$.

In addition to the definition given in Section \ref{sec:3.3}, chirality of spin waves can be defined by the following broken symmetry of the dispersion (\ref{ReOmegaq}):
\begin{equation}\label{ChirOmega}
    \omega_{q}^{B_{\rm ext}}=\omega_{-q}^{-B_{\rm ext}}\quad\text{and}\quad\omega_{q}^{B_{\rm ext}}\neq\omega_{-q}^{B_{\rm ext}}.
\end{equation}
The property (\ref{ChirOmega}) is illustrated in Figs. \ref{fig14}a-b, which present the energy and linewidth dispersions of the type of spin wave shown in Fig. \ref{fig13} for both directions of the magnetic field, but in a sample meeting the above condition. Since the linewidth of the Raman line yields the damping rate $\eta_q$, Figs.~\ref{fig14}a-b shows strikingly that both the spin-wave energy and damping rate exhibit the same chirality:  they are invariant under simultaneous inversion of the directions of the magnetic field and of the wavevector. Moreover one can extract a momentum shift $q_s\simeq 1.5 \mu{\rm m}^{-1}$ which shifts simultaneously the extremal energy and damping from $q=0$.

When changing the in-plane angle $\varphi$ of $\bfq$ for which the dispersions are probed, a modulation of $\bfq_s$ with $\varphi$ appears, as shown in Fig.~\ref{fig14}c. The $\pi$-periodicity of the $q_s(\varphi)$ modulation is in complete agreement with the $C_{2v}$ in-plane symmetry of the SOC arising from the superposition of the Rashba and Dresselhaus contributions (see Section \ref{sec:2.4}) and confirms the SOC origin of the observed chirality.

Chirality in \SWd energy dispersions and chiral damping have been observed in Fe monolayers \cite{Zakeri2012}. Chiral damping dispersions have been observed in Pt/Co/Ni films \cite{Kai2015}. However, Eqs.~(\ref{ReOmegaq}) and (\ref{ImOmegaq}) show a universal linear relation between damping rate and angular frequency of the spin wave, independent of SOC, which reads:
\begin{equation}\label{Lin}
 \eta_q = \tilde \eta_0 - \frac{2m^*}{\hbar }\frac{\eta_2}{\sws}\omega_q,
\end{equation}
where $\tilde \eta_0 = \eta_0 + 2mZ/\hbar^2 \sws$. This universal linear behavior survives to the presence of SOC as demonstrated in Figs.~\ref{fig14}d-e where the linewidth is plotted as a function of energy for $B_{\rm ext}=+1{\rm T}$ and $B_{\rm ext}=+2{\rm T}$ and various in-plane angles, which means various strengths of SOC. The chirality and anisotropy do not appear anymore: $+\bfx$ and $-\bfx$ waves, for every $\varphi$, fall on the same line. This linear relation of Figs.~\ref{fig14}d-e was not found in Ref. \cite{Kai2015}. This underlines the particular physics of chiral spin-waves in 2DEL, which is due to the underlying symmetries of SOC. Indeed, the SOC of the Hamiltonian (\ref{H}) can be removed by a unitary transformation \cite{Perez2016}. In the transformed reference frame, position and phase of the spin motion are locked by the quantity $\bfq_0 \cdot \bfr_i$ where $\bfr_i$ is the electron-spin position and

\begin{equation}\label{q0}
\bfq_{0}=\frac{2m^{\ast }}{\hbar^2 }[\left( \alpha + \beta \sin 2\varphi\right)\bfx
+\beta \cos 2\varphi\bfz ]
\end{equation}
is a SOC-dependent constant wavevector. As a consequence, the dispersions of Eqs. (\ref{ReOmegaq}) and (\ref{ImOmegaq}) obtained without SOC, are, with SOC, simply shifted in $\bfq$-space by $\bfq_{0}$, and the complex angular frequency becomes:
\begin{equation}\label{hom2q0}
\hbar\tilde\omega\sw^{\rm SO}(\bfq) =  Z -\sws\frac{\hbar^2 }{2m^*}|\bfq +\bfq_0|^2+i\hbar\eta_{\bfq +\bfq_0} \:.
\end{equation}
This introduces in $\omega_{q}$ a modulation term: $-\sws q ( \alpha + \beta \sin 2\varphi)$ fully compatible with Fig.~\ref{fig14}c and $q_s=-\bfq_0 \cdot \bfx$.
We point out that Eq. (\ref{hom2q0}) is correct to leading (first) order in the SOC strengths $\alpha$ and $\beta$.

\subsection{Larmor's mode in the presence of SOC}  \label{sec:4.4}

In the presence of SOC, Larmor's mode frequency is no longer the bare Zeeman energy $Z$, but is corrected by anisotropic contributions that are
of second order in the SOC strengths $\alpha$ and $\beta$ \cite{Karimi2017}. Equation (\ref{hom2q0}) seems to suggest that
\begin{eqnarray}\label{E0approx}
\hbar \omega\sw^{\rm SO}(q=0) &\approx&  Z - 2m^*\sws(\alpha^2+\beta^2 \nonumber\\
&&{} +2\alpha\beta\sin2\varphi).
\end{eqnarray}
As the spin-wave stiffness $\sws$ contains Coulombic contributions, SOC induces a breaking of the Larmor's theorem (\ref{Larmor}). Note that, since $q=0$, changing the $\varphi$ angle has to be understood as tuning the precession direction of the spins with respect to the crystalline axis. The Rashba SOC is isotropic and simply shifts the frequency, so that the additional presence of Dresselhaus SOC is responsible for this anisotropy.

\begin{figure}
\centering
\includegraphics[width=\columnwidth]{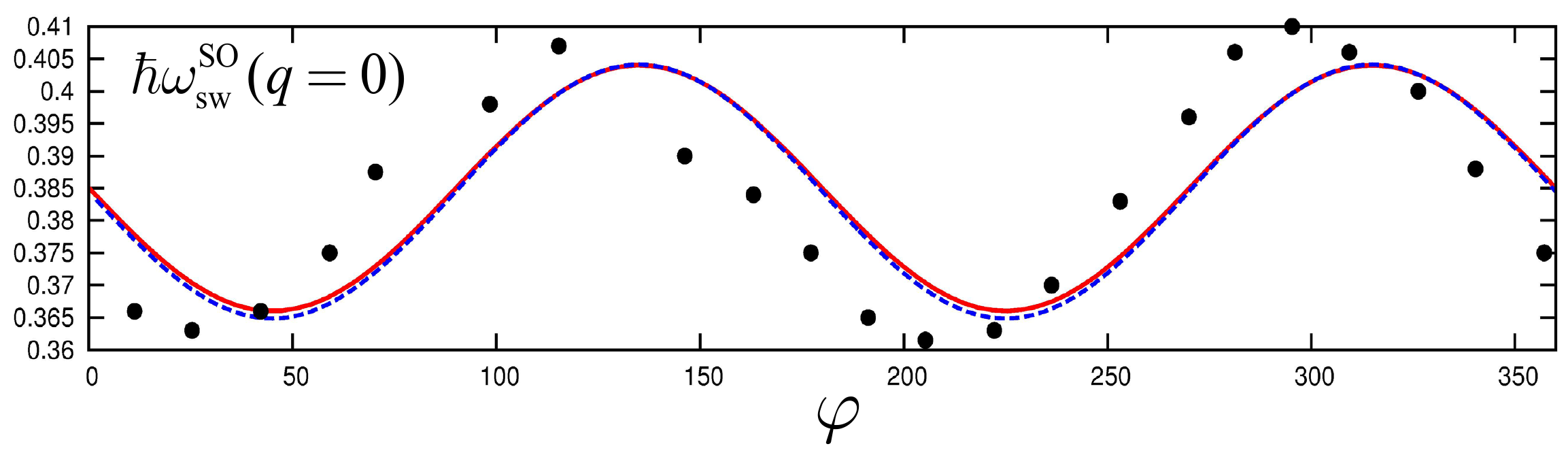}
\caption{Larmor's mode energy $\hbar \omega\sw^{\rm SO}(q=0)$ as a function of angle $\varphi$. Dots: experimental
data. Lines: theoretical results using Eq. (\ref{Larmor0}) (dashed blue) and fully numerical linear-response TDDFT-ALDA (red).}
\label{fig15}
\end{figure}

The breaking of the Larmor's theorem has been studied carefully and evidenced through this anisotropic correction by ERRS in Ref. \cite{Karimi2017}. Results are reproduced in  Fig.~\ref{fig15}. It is shown that the experimental data are reproduced by TDDFT linear-response theory if SOC is taken into account beyond first order.
This yields the following result for the Larmor's mode frequency, correct to second order in SOC:
\begin{eqnarray}\label{Larmor0}
    \hbar \omega\sw^{\rm SO}(q=0) &=& Z +\frac{2\pi n_{\rm 2D}}{Z^* \bar f_{\rm xc}}[(\alpha^2+\beta^2)(3 \bar f_{\rm xc}+2) \nonumber\\
    &+&2\alpha\beta\sin2\varphi(\bar f_{\rm xc}+2)].
\end{eqnarray}
Here, $\bar f_{\rm xc}$ can be calculated using the ALDA xc kernel averaged over the lowest subband envelope function, or fitted using
$\bar f_{\rm xc}=Z/Z^*-1<0$. Equation (\ref{Larmor0}) has the same $\sin 2\varphi$ anisotropy as the approximate Eq. (\ref{E0approx}), but with slightly
different coefficients.

We conclude this section by pointing out that expressions similar to (\ref{hom2q0})--(\ref{Larmor0}), including situations where $\bfq$ is not
perpendicular to $\bfB_{\rm ext}$,  were obtained in Ref. \cite{Maiti2017}  using diagrammatic techniques.

\begin{figure*}
\centering
\includegraphics[width=\textwidth]{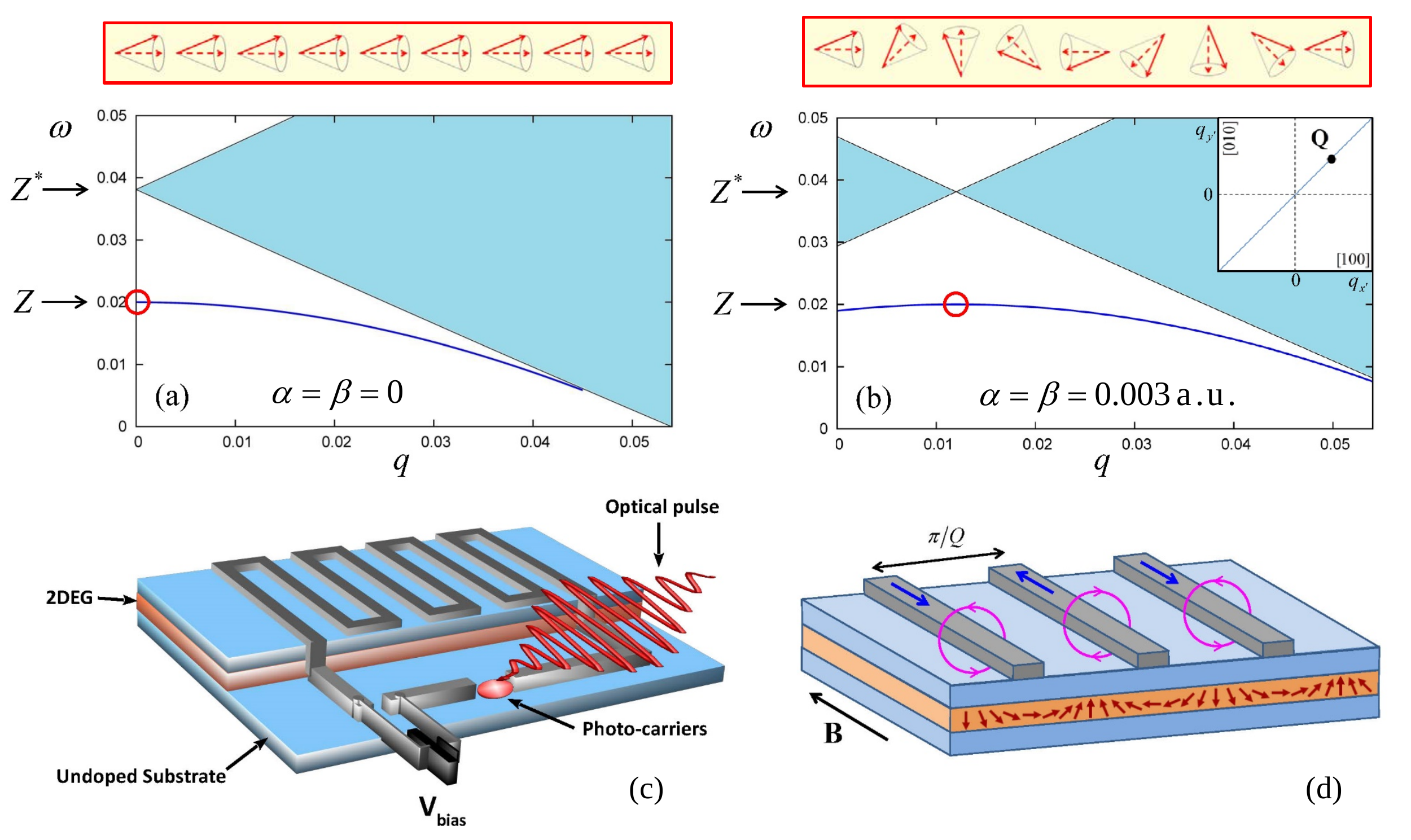}
\caption{(a) Spin-wave dispersion in the absence of SOC, illustrating the usual Larmor mode (circled in red), in which all spins precess
about the direction of the in-plane magnetic field. (b) Spin-wave dispersion in the presence of SOC, with $\alpha=\beta$.
The spin-wave dispersion along the [110] direction is the same as for the system without SOC, but shifted by a wavevector $\bfQ=4\alpha$.
The spin-helix Larmor mode (circled in red) is a standing-wave mode, precessing about the spin-helix texture.
(c) Proposed experimental design for the optical excitation of the spin-helix Larmor mode, using a photoconductive antenna.
(d) close-up view of the metal stripes on top of the 2DEL, showing a proposal for detection of the mode by the induced alternating currents
triggered by the standing spin wave. Adapted from Ref. \cite{Karimi2018}. }
\label{fig16}
\end{figure*}

\subsection{Spin-Helix Larmor mode}  \label{sec:4.5}

We now consider a very special case in which exact results can be proved to all orders in SOC, namely, the case of
a persistent spin helix
\cite{Schliemann2017,Koralek2009,Schliemann2003,Bernevig2006,Walser2012,Schonhuber2014,Sasaki2014,Fu2017}.
The spin helix arises in a 2DEL in which the Rashba and Dresselhaus coupling strengths are equal, i.e., $\alpha=\beta$.
We here limit the discussion to a 2DEL embedded in a zincblende quantum well grown along the [001] direction:
SU(2) symmetry is then partially restored, and a helical spin texture can be sustained along the  [110] direction.
This property is protected against decoherence from spin-independent disorder scattering and Coulomb interactions \cite{Bernevig2006},
and leads to the experimentally observed extraordinarily long lifetimes of spin packet excitations \cite{Koralek2009,Sasaki2014}.

Without any applied magnetic field, the spin helix states are exact single-particle eigenstates in the 2DEL; this is caused
by a degeneracy of the two branches of the energy dispersion (see the left panel of Fig. \ref{fig5}) of the form
$E_{+,\bfk+\bfQ} = E_{-,\bfk}$, where $\bfQ = 4\alpha \hat e_{[110]}$. A superposition of any two degenerate states
on the two branches then has a helical structure, see Ref. \cite{Bernevig2006}.

If a magnetic field is applied in the plane of the 2DEL, perpendicular to the [110] direction, then this degeneracy is lifted (see
right panel of Fig. \ref{fig5}). Instead, the spin helix becomes a nonequilibrium feature, where spin-flip single-particle excitations
give rise to propagating spin helices \cite{Walser2012}.

If we now include collective effects due to Coulomb interactions,
it becomes possible to prove an exact many-body result for spin waves, which we call the {\rm spin-helix Larmor mode} \cite{Karimi2018}: if,
in a system with $\alpha=\beta$, the spin wave has wavevector $\bfQ$ commensurate with the spin-helix texture, all Coulomb
interaction contributions drop out, and the spin-wave frequency is given by the bare Zeeman energy:
\begin{equation}
\omega_{\rm sw}^{\alpha=\beta\ne 0}(\bfQ) = \omega_{\rm sw}^{\alpha=\beta=0}(0) = Z \:.
\end{equation}
In other words, Larmor's standing-wave precessional mode now occurs with a finite wavevector.

Panels (a) and (b) of Fig. \ref{fig16} compare the spin-wave dispersions with and without SOC, and the cartoons on top of the panels illustrate the spin dynamics at
the special points marked by the red circles. The usual Larmor mode at $\bfq=0$ is characterized by a collective precession about a spatially fixed axis,
whereas the spin-helix Larmor mode at $\bfq=\bfQ$ is characterized by a collective precession about an axis that rotates in space, but is not itself propagating.
Both Larmor modes are undamped in the absence of any extrinsic mechanisms.

Panels (c) and (d) of Fig. \ref{fig16} show a proposal for direct optical excitation of the spin-helix Larmor mode via a photo-conductive antenna,
using the same antenna for detection of the alternating currents associated with the magnetic fields induced by the standing spin wave.

\begin{figure}[]
\centering
\includegraphics[width=\columnwidth]{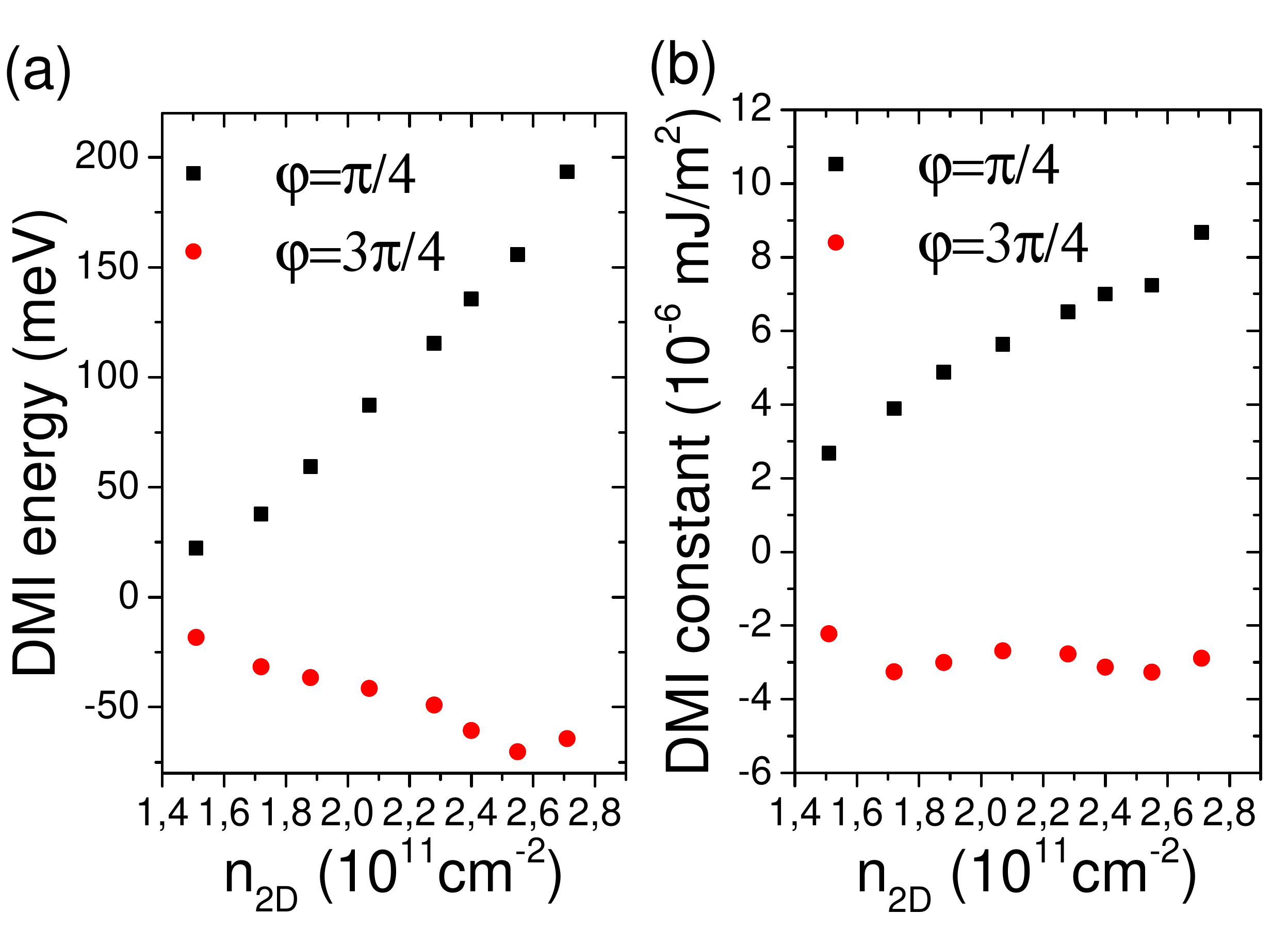}
\caption{(a) DMI energy as function of 2DEL density gated by illumination. As the chiral shift depends sinusoidally on the in-plane direction, the two extremal directions ($\varphi=\pi/4$ and $\varphi=3\pi/4$) are shown. The DMI energy is to be compared with $0.9 \,{\rm meV}$ found in Ref. \cite{Zakeri2012}. (b) DMI constant, to be compared with 0.44~mJ/m$^2$ found in Ref. \cite{Kai2015}. }
\label{fig17}
\end{figure}

\subsection{Comparison with DMI} \label{sec:4.6}

Chiral spin waves meeting condition (\ref{ChirOmega}) have been observed in ferromagnets \cite{Zakeri2012,Zakeri2010}.  An extensive body of
literature has been devoted to describe such spin waves within the Dzyaloshinskii-Moriya interaction (DMI), an asymmetric Heisenberg-type exchange:

\begin{equation}\label{DMI}
    \h_{\rm Ferro}=\sum_{<ij>}J_{ij}\hbS_i\cdot\hbS_j+\sum_{<ij>}D_{ij}\hbS_i\times\hbS_j .
\end{equation}
In most systems the DMI energy terms $D_{ij}$ remain empirical parameters with a magnitude of a few percent of the exchange energy $J_{ij}$ \cite{Nembach2015,Dmitrienko2016,Lee2016}. The microscopic origin of the DMI term can be SOC \cite{Udvardi2009,Costa2010}. The DMI approach is perfectly well suited for spins strongly or weakly localized. However, for delocalized spins in a Galilean invariant system, such as a 2DEL, the main subject of this review, one can show that the DMI interaction cannot reproduce the physics outlined in Fig.~\ref{fig14}.
Indeed, transforming Eq. (\ref{DMI}) into a continuous system of spins, coupled by a DMI step function $D(r)$, is equivalent to an isotropic $D_{ij}$ for $r<a_B^*$ (inside the Pauli hole) and zero elsewhere. Inserting the DMI part of Eq. (\ref{DMI}) into Eq. (\ref{Sqmot}), and comparing with the $q$-linear term in Eq. (\ref{hom2q0}), we can deduce the following, $\varphi$-dependent DMI energy (the coupling between neighboring spins) \cite{Zakeri2012}:

\begin{equation}\label{DMIphi1}
    D_\varphi^E=\frac{2}{3}\,\frac{D_{ij}}{r_s^2}=\sws\frac{4}{a_B^*\zeta}( \alpha + \beta \sin 2\varphi),
\end{equation}
where $a_B^*$ is the material Bohr radius (see Table \ref{Tabrs}) and $\zeta$ is the spin-polarization degree. Alternatively, we can deduce a DMI constant (the DMI energy times magnetization) \cite{Kai2015}:
\begin{equation}\label{DMIphi2}
    D_\varphi^C=\sws\frac{n_{\rm 2D}\zeta}{2w}( \alpha + \beta \sin 2\varphi),
\end{equation}
where $w$ is the quantum well thickness. As shown in Fig.~\ref{fig17}, we find DMI energies 20 times larger than what was found by Zakeri \emph{et al.} \cite{Zakeri2012}
and DMI constants 5 orders of magnitude below what was found by Di \emph{et al.} \cite{Kai2015}. The DMI framework, which is based on the fundamental assumption of localized spins, is obviously inconsistent with the Galilean invariance of the 2DEL.

\section{Conclusion and perspectives}  \label{sec:5}

In this review, we have argued that the 2DEL---a model system which has been the subject of intense scrutiny for
many decades---still holds many surprises. We have focused on collective spin excitations of the 2DEL,
both of a longitudinal (spin conserving) and transverse (spin-flip) nature.
The presence of SOC (Rashba and Dresselhaus), in-plane magnetic fields, and magnetic impurities at first seems to complicate matters enormously; indeed, the single-particle properties exhibit increasing degrees of complexity upon the addition of new features in the Hamiltonian.

However---and surprisingly---when Cou\-lomb many-body effects enter the game,  collective behavior emerges which, in the end, leads to dramatic simplifications. Spin plasmons and spin waves behave, in many ways, as macroscopic quantum object, subject to precession or Zeeman effects, similar to what is seen in simple one- or few-body quantum systems.
Thanks to Coulomb interaction, the collective modes are protected against dephasing due to SOC.
Through a unitary transformation, exact many-body results such as Larmor's theorem find their counterparts in the presence of SOC. However,
things don't always get simpler: Coulomb many-body effects also provide intrinsic sources of dissipation for the collective modes, most notably via the frictional forces due to the SCD. There are a number of issues related to the SCD that remain to be resolved, as we will now discuss.

The longitudinal SCD is nonzero even for $q=0$: in this case the corresponding spin-transresistivity in high-mobility systems has been predicted to be comparable or larger than the Drude resistivity \cite{Damico2001,Damico2003}, and to be stronger in low-dimensional systems \cite{Flensberg2001,Damico2003}.
Related theoretical predictions for significant reduction of spin diffusion due to longitudinal SCD \cite{Damico2001} were initially confirmed experimentally in \cite{Weber2005} and more recently in \cite{Yang2012} and \cite{Walser2012}. The longitudinal SCD is predicted to not affect the spin mobility if the scattering times for the
two spin components are similar \cite{Damico2001,Damico2002}; this was confirmed experimentally for spin propagation in a GaAs 2DEL \cite{Yang2012}. However, the SCD should affect the spin mobility for spin-dependent scattering \cite{Damico2002},  which is relevant for spin packets and spin-wave propagation. Experimental confirmation of this effect is still awaiting.

Experimental results on the dissipation of intersubband spin plasmons \cite{Baboux2012} led to the development of a fully non-homogeneous theoretical treatment of longitudinal SCD \cite{Damico2013}. While this predicts the SCD to be a sizeable source of dissipation for collective spin modes (e.g. up to 20\% of the linewidth measured in \cite{Baboux2012}), a direct observation disentangling SCD from extrinsic sources of dissipation is still lacking.

The transverse SCD  is predicted to contribute to the Gilbert damping  of a magnetized 2DEL \cite{Hankiewicz2008}, and hence to the intrinsic dissipation of transverse spin waves. The effect vanishes as $q\to0$  and so far has eluded direct experimental verification \cite{Gomez2010}.

In the presence of SOC, both longitudinal and transverse SCD contribute to the Gilbert damping  of a spin-polarized 2DEL \cite{Hankiewicz2007}, both for homogenous and modulated electronic systems \cite{Hankiewicz2007}. Indeed, SCD is enhanced by the presence of SOC \cite{Tse2007}, as SOC increases scattering between different-spin populations: this SOC-enhanced Coulomb-induced dissipation remains nonzero at $q=0$, both for longitudinal and transverse SCD. For weak SOC, the strength of the damping is proportional to the square of the SOC coupling renormalized by the Fermi energy \cite{Hankiewicz2007,Tse2007,Maiti2015a}. This dependence, and the related damping of collective spin modes \cite{Hankiewicz2007,Maiti2015a}, have not yet been confirmed experimentally.

Coulomb interactions affect the diffusion of spin packets (and hence of spin waves) both through the spin stiffness and through SCD \cite{Damico2001,Damico2002}.  As a consequence, during a paramagnetic to ferromagnetic transition, the spin diffusion is predicted to undergo large variations, including, in certain cases, vanishing \cite{Damico2001}. Similar predictions stand for related quantities \cite{Damico2010}, and hold both for ordinary electron liquids and for semiconductors doped with magnetic impurities. These consequences of Coulomb interactions remain to be experimentally explored.

We feel that cross-collaboration between theory and experiments is fundamental to further this field. For example in Ref. \cite{Baboux2012}, excellent agreement between theory and experiments for the linewidth modulation of intersubband spin-plasmons was achieved; however more experimental and theoretical studies are necessary to achieve full consistency between first-principles predictions by the DP mechanism  and experimental results, in particular, time domain measurements (with TSG) of the chiral spin-waves decay as a function of the in-plane wavevector $q$ \cite{Weber2005} and measurements of the Larmor's mode lifetime at finite $q$. A comparison with measurements of the conductivity would also be enlightening.

In general, improving first-principles treatments of dissipation within (TD)DFT is an enduring challenge for the DFT community, as it requires inclusion of non-Markovian processes (memory) within the formalism: contrary to the formalism reviewed in this article, the most widely used TDDFT approximation, adiabatic-LDA, does not include memory effects, so that in adiabatic-LDA dissipation has to be included 'by-hand'. Precision experimental linewidth data serves as an important
benchmark for the developers of new, nonadiabatic xc functionals in TDDFT \cite{Ullrich2001,Ullrich2002a}.

New materials platforms  that support low-dimensional quantum liquids  continue to emerge. Here, we have focused on 2DELs in semiconductor quantum wells, which have been well characterized for decades. However, 2D materials such as the graphene family and beyond-graphene materials offer innumerable opportunities to explore collective electronic phenomena; features such as linear dispersions (Dirac electrons), topological invariants, or the interplay between different valleys remain to be studied.  We believe that the concepts exposed here are largely transferable to the emerging 2D systems such as dichalcogenides monolayers \cite{Butler2013,Novoselov2016}.  These materials exhibit very large SOC and strong Coulomb coupling strength. Ideal would be to find a system were these effects survive at room temperature. This requires a scaling of the strength of all the protagonists by at least a factor of ten. If such a material emerges, devices can be made were the interplay between SOC and Coulomb interactions can be utilized to build a spin-wave based transistor like in Ref. \cite{Kajiwara2010}. As in Ref. \cite{Perez2016}, it would then be easy to manipulate the group velocity of a spin wave, or tune its direction of propagation, or modulate its phase, all together with gated electrodes acting on the magnetic layer.

In conclusion, we are confident that the study of collective spin dynamics in low-dimensional electronic systems will continue to lead to new discoveries in basic physics, and will set the stage for new applications in (quantum) technologies.

\section*{Acknowledgments}

F.P. acknowledges support from the Fondation CFM, C'NANO IDF and ANR.
C. A. U. acknowledges financial support by DOE Grants DE-FG02-05ER46213 and DE-SC0019109.
I. D. acknowledges hospitality and partial financial support by the International Institute of Physics, Federal University of Rio Grande do Norte, Natal, Brazil.

\newpage

\bibliographystyle{unsrt}
\bibliography{CU_IDA_FP_refs}

\end{document}